\documentclass[aip,reprint,author-year,nofootinbib]{revtex4-1}

\usepackage{graphicx}
\usepackage{microtype}
\usepackage{amsmath,amssymb}             
\usepackage{stmaryrd}					
\usepackage[normalem]{ulem}
\usepackage{color}
\usepackage[dvipsnames]{xcolor}
\usepackage{color,hyperref}
\definecolor{darkblue}{rgb}{0.0,0.0,0.4}
\definecolor{red}{rgb}{1.0,0.0,0.0}
\definecolor{green}{rgb}{0.0,0.5,0.0}
\hypersetup{colorlinks,breaklinks,
            linkcolor=darkblue,urlcolor=darkblue,
            anchorcolor=darkblue,citecolor=RoyalBlue}

\usepackage{gensymb} 
\usepackage[mathscr]{euscript}
\usepackage{upgreek}
\DeclareFontFamily{OT1}{pzc}{}
\DeclareFontShape{OT1}{pzc}{m}{it}{<-> s * [1.10] pzcmi7t}{}
\DeclareMathAlphabet{\mathpzc}{OT1}{pzc}{m}{it}
\usepackage{stmaryrd}

\def\apjl{Astrophys. J. Lett.}
\def\apj{Astrophys. J.}
\def\aj{Astronom. J.}

\def\apjs{Astrophys. J. Supp.}

\def\mnras{Mon. Not. R. Astron. Soc.}

\def\prd{Phys. Rev. D}

\def\physrep{Physics Reports}
\def\nat{Nature}

\def\aap{Astron. \& Astrophys.}

\def\jcap{Journal of Cosmology and Astroparticle Physics}

\newcommand{\comment}[1]{}

\defcitealias{Jasche2015BORGSDSS}{\textsc{borg sdss}--inference}
\defcitealias{Leclercq2015ST}{\textsc{borg sdss}--{\tweb}}
\newcommand{\borg}{\textsc{borg}}
\newcommand{\cola}{\textsc{cola}}
\newcommand{\tweb}{\textsc{T-web}}
\newcommand{\diva}{\textsc{diva}}
\newcommand{\lich}{\textsc{lich}}
\newcommand{\origami}{\textsc{origami}}

\begin{document}


\title{The phase-space structure of nearby dark matter as constrained by the SDSS}


\author{Florent Leclercq}
\email{florent.leclercq@polytechnique.org}
\thanks{\href{http://icg.port.ac.uk/~leclercq/}{http://icg.port.ac.uk/$\sim$leclercq/}}
\affiliation{Institute of Cosmology and Gravitation (ICG), University of Portsmouth,\\ Dennis Sciama Building, Burnaby Road, Portsmouth PO1 3FX, United Kingdom}

\author{Jens Jasche}
\affiliation{Excellence Cluster Universe, Technische Universit\"at M\"unchen,\\ Boltzmannstrasse 2, D-85748 Garching, Germany}

\author{Guilhem Lavaux}
\affiliation{Institut d'Astrophysique de Paris (IAP), UMR 7095, CNRS -- UPMC Universit\'e Paris 6, Sorbonne Universit\'es, 98bis boulevard Arago, F-75014 Paris, France}
\affiliation{Institut Lagrange de Paris (ILP), Sorbonne Universit\'es,\\ 98bis boulevard Arago, F-75014 Paris, France}

\author{Benjamin Wandelt}
\affiliation{Institut d'Astrophysique de Paris (IAP), UMR 7095, CNRS -- UPMC Universit\'e Paris 6, Sorbonne Universit\'es, 98bis boulevard Arago, F-75014 Paris, France}
\affiliation{Institut Lagrange de Paris (ILP), Sorbonne Universit\'es,\\ 98bis boulevard Arago, F-75014 Paris, France}
\affiliation{Department of Physics, University of Illinois at Urbana-Champaign,\\ 1110 West Green Street, Urbana, IL~61801, USA}
\affiliation{Department of Astronomy, University of Illinois at Urbana-Champaign,\\ 1002 West Green Street, Urbana, IL~61801, USA}

\author{Will Percival}
\affiliation{Institute of Cosmology and Gravitation (ICG), University of Portsmouth,\\ Dennis Sciama Building, Burnaby Road, Portsmouth PO1 3FX, United Kingdom}


\date{\today}

\begin{abstract}
\noindent Previous studies using numerical simulations have demonstrated that the shape of the cosmic web can be described by studying the Lagrangian displacement field. We extend these analyses, showing that it is now possible to perform a Lagrangian description of cosmic structure in the nearby Universe based on large-scale structure observations. Building upon recent Bayesian large-scale inference of initial conditions, we present a cosmographic analysis of the dark matter distribution and its evolution, referred to as the dark matter phase-space sheet, in the nearby universe as probed by the Sloan Digital Sky Survey main galaxy sample. We consider its stretchings and foldings using a tetrahedral tessellation of the Lagrangian lattice. The method provides extremely accurate estimates of nearby density and velocity fields, even in regions of low galaxy density. It also measures the number of matter streams, and the deformation and parity reversals of fluid elements, which were previously thought inaccessible using observations. We illustrate the approach by showing the phase-space structure of known objects of the nearby Universe such as the Sloan Great Wall, the Coma cluster and the Bo\"{o}tes void. We dissect cosmic structures into four distinct components (voids, sheets, filaments, and clusters), using the Lagrangian classifiers {\diva}, {\origami}, and a new scheme which we introduce and call {\lich}. Because these classifiers use information other than the sheer local density, identified structures explicitly carry physical information about their formation history. Accessing the phase-space structure of dark matter in galaxy surveys opens the way for new confrontations of observational data and theoretical models. We have made our data products publicly available.
\end{abstract}


\maketitle



\section{Introduction}

The accurate description of large-scale structure formation, from which the cosmic web originates \citep{Bond1996}, is one of the major goals of modern cosmology. Traditionally, the Lagrangian and Eulerian descriptions of a discretized fluid of collisionless dark matter are comparatively discussed. In the Eulerian picture, quantities are tracked at fixed positions as particles move. In the Lagrangian framework, the motion and evolution of individual elements is followed. Therefore, particles can be thought of not only as mass tracers, but also as elements of a three-dimensional dark matter sheet, which stretches, folds and distorts in a six-dimensional velocity-position phase space.

Using cosmological simulations, several recent papers have demonstrated the power of working within the dark matter sheet. Tessellations of the initial Lagrangian space allow the identification of stream crossings and improve density measurements \citep{Abel2012,Shandarin2012}, provide accurate estimates of cosmic velocity fields \citep{HahnAnguloAbel2015}, and yield a new approach for simulating dark matter \citep{Hahn2013}. In addition, the study of caustics \citep{Arnold1982,White2009,VogelsbergerWhite2011,Hidding2014,Feldbrugge2014}, the two-dimensional edges along which the dark matter sheet folds, brings valuable insights into structure formation \citep{Neyrinck2012}. Strong observational motivations exist for a Lagrangian understanding of the cosmic web. These include studying the dependence of galaxy properties on their evolving environment \citep[e.g.][]{Blanton2005a,Libeskind2015}, probing the effect of the dynamic large-scale structure on the cosmic microwave background \citep[e.g.][]{LavauxAfshordiHudson2013,Planck2013ISW}, testing the standard general-relativistic picture of gravitational instability \citep{Falck2015a}, and identifying the most promising regions to observe a possible dark matter annihilation signal \citep[e.g.][]{Gao2012}.

Unfortunately, the Lagrangian picture cannot be readily applied to observational data. Difficulties are of two sorts: (i) the need for reconstructions of the dark matter density from galaxies, accounting for typical observational effects such as biases, survey mask and selection functions; and (ii) the need for corresponding initial conditions, which requires the inclusion of a physical picture of structure formation into the data model. For these reasons, most cosmic web analyses have been limited so far to simulations \citep[see e.g.][for a recent study]{Cautun2014}. However, recently, \citet{Jasche2010a} and \citet[][hereafter \citetalias{Leclercq2015ST}]{Leclercq2015ST} addressed the first issue. Using real data, they presented Eulerian classifications of cosmic environments in constrained realizations of the large-scale structure. Importantly, they demonstrated capability of propagating observational uncertainties to cosmic web classification. Taking advantage of the inclusion of a physical model within the inference process, \citetalias{Leclercq2015ST} also presented the first probabilistic analysis of proto-structures present in the initial density field. However, a complete investigation of the problem, including a probabilistic description of phase-space properties of dark matter based on real data, has not yet been performed in the literature.

This paper describes the first analysis of the Lagrangian dark matter sheet in the nearby Universe, as probed by galaxies of the northern galactic cap of the Sloan Digital Sky Survey (SDSS) main sample. Capitalizing on the full phase-space information, we produce highly-detailed and accurate maps of cosmic density and velocity fields, even in regions only poorly sampled by galaxies. We also analyze additional information specific to collisionless fluids, which was so far inaccessible in observations: the number of matter streams, the deformation and parity of fluid elements, and the presence of caustics. We present the first application of the Lagrangian cosmic web classifiers {\diva} and {\origami} to the real Universe, as well as a new algorithm, {\lich}, which we design to take into account the heterogeneous (potential and vortical) nature of flows. 

This chrono-cosmography project builds upon the inference of the past and present cosmic structure in the SDSS \citep[][hereafter \citetalias{Jasche2015BORGSDSS}]{Jasche2015BORGSDSS}, performed using the {\borg} algorithm \citep{Jasche2013BORG}. It also relies on a large set of constrained realizations of the Universe, produced using the {\cola} method \citep{Tassev2013}. These necessarily introduce some degree of extrapolation with respect to the original inference results, since they contain aspects of the physical model that have not been constrained by the data (in particular, in this work, vortical matter flows). However, these extrapolations are not arbitrary, but are methodical predictions that need to be consistent with the observational constraints. Indeed, a complex forward model naturally introduces correlations between its constrained and unconstrained aspects, which may yield reliable predictions for the physics that has not been captured by the simpler data model. Our methodology therefore predicts the properties of the Lagrangian dark matter sheet based on fusing prior information from simulations and data constraints. Such an approach has been previously demonstrated to accurately recover the properties of voids at the level of the dark matter distribution \citep{Leclercq2015DMVOIDS}. Importantly, all of our predictions are fully probabilistic, which means that their degree of speculativeness is controlled and can be checked against unconstrained simulations. From a Bayesian perspective, all the results presented in this work can then be used as prior information for updating our knowledge in light of additional observations, or for making optimal decisions in the presence of uncertainty \citep[see][]{Leclercq2015DT}.

The structure of this paper is as follows. We start by describing our tools for analyzing the dark matter sheet in section \ref{sec:Description of the Lagrangian sheet}. This includes a review of the Lagrangian displacement field, the description of the existing Lagrangian classifiers {\diva} and {\origami} and the introduction of a new one that we call {\lich}, and the definition of estimators based on tessellating the Lagrangian particle grid. In section \ref{sec:Lagrangian analysis of the SDSS volume}, we then infer, characterize and classify the SDSS volume in Lagrangian space. Finally, in section \ref{sec:Consequences in Eulerian coordinates}, we show the consequences of these results in Eulerian coordinates and demonstrate that resulting maps of the redshift-zero cosmic web have an interesting level of predictive power. We summarize our study and conclude in section \ref{sec:Summary and conclusions}.

\section{Description of the Lagrangian sheet}
\label{sec:Description of the Lagrangian sheet}

\subsection{The Lagrangian displacement field}

The fundamental object studied in this paper is the displacement field $\Psi(\textbf{q})$, which maps the initial (Lagrangian) position of particles $\textbf{q}$ to their final (Eulerian) position $\textbf{x}(\textbf{q})$ \citep[see e.g.][for a classic review]{Bernardeau2002}:
\begin{equation}
\textbf{x}(\textbf{q}) \equiv \textbf{q} + \Psi(\textbf{q}).
\label{eq:mapping}
\end{equation}

As any smooth vector field, it is possible to split $\Psi$ into a scalar and a rotational part \citep[the Helmholtz decomposition,][]{Chan2014},
\begin{equation}
\Psi(\textbf{q}) = \nabla_\textbf{q} \Phi_\Psi + \nabla_\textbf{q} \times \textbf{A}_\Psi,
\end{equation}
where $\Phi_\Psi$ is the scalar potential and $\textbf{A}_\Psi$ is the vector potential, which obey the Poisson equations:
\begin{eqnarray}
\Delta \Phi_\Psi(\textbf{q}) & = & \nabla_\textbf{q} \cdot \Psi(\textbf{q}) \equiv \psi(\textbf{q}), \\
\boldsymbol{\Delta} \textbf{A}_\Psi(\textbf{q}) & = &- \nabla_\textbf{q} \times \Psi(\textbf{q}) \equiv \textbf{w}(\textbf{q}).
\end{eqnarray}
From analytic approaches and numerical simulations, it has been known for some time that the displacement field is almost fully potential; in particular, $\psi$ contains all the information up to second order in Lagrangian perturbation theory (2LPT). In this paper, we estimate $\psi(\textbf{q})$ and $\textbf{w}(\textbf{q})$ by computing respectively $\mathcal{F}^{-1}[ -\mathrm{i} \textbf{k} \cdot \mathcal{F}[ \Psi ]]$ and $\mathcal{F}^{-1}[ \mathrm{i} \textbf{k} \times \mathcal{F}[ \Psi ]]$, where $\mathcal{F}$ is a fast Fourier transform (FFT) on the Lagrangian mesh on which the displacement field is defined. For later use, we define $\Upsilon(\textbf{q}) \equiv \lVert\nabla_\textbf{q}~\times~\textbf{A}_\Psi\rVert$.

We denote by $J(\textbf{q})$ the Jacobian determinant of the transformation between Lagrangian and Eulerian coordinates, 
\begin{equation}
J(\textbf{q}) \equiv \det \left[ \frac{\partial \textbf{x}}{\partial \textbf{q}} \right] = \det \mathscr{D} = \det ( \mathscr{I} + \mathscr{R} ),
\end{equation}
where the deformation tensor $\mathscr{D}_{\ell m}$ can be written as the identity tensor $\mathscr{I}_{\ell m} \equiv \updelta_\mathrm{K}^{\ell,m}$ plus the shear of the displacement field $\mathscr{R}_{\ell m} \equiv \partial \Psi_\ell /\partial \textbf{q}_m$. The characteristic polynomial of $\mathscr{R}$ is written as
\begin{equation}
\mathcal{P}_\mathscr{R}(\lambda) \equiv \det (\mathscr{R} - \lambda \mathscr{I}) = - \textstyle{\prod\limits_\ell} (\lambda-\lambda_\ell).
\label{eq:characteristic_polynomial_R}
\end{equation}
where the $\lambda_i$ are the three eigenvalues (real or complex) of $\mathscr{R}$ (i.e. the solutions of $\mathcal{P}_\mathscr{R}(\lambda)=0$). The Jacobian is then $J(\textbf{q}) = (1+\lambda_1(\textbf{q}))(1+\lambda_2(\textbf{q}))(1+\lambda_3(\textbf{q}))$. 

The characteristic equation $\det (\mathscr{R} - \lambda \mathscr{I}) = 0$ can be written 
\begin{equation}
\lambda^3 + s_1 \lambda^2+ s_2 \lambda +s_3 = 0,
\end{equation}
where we call the $s_i$ the \textit{Lagrangian invariants} \citep[in analogy to the \textit{rotational invariants} introduced by][who use the shear of the Eulerian velocity field instead of the Lagrangian displacement field]{Wang2014}. They are given in terms of the eigenvalues by \citep{Chong1990}
\begin{eqnarray}
s_1(\textbf{q}) & = & - \textstyle{\sum\limits_\ell} \lambda_\ell(\textbf{q}), \label{eq:laginv_formulae_1}\\
s_2(\textbf{q}) & = & \textstyle{\sum\limits_{\ell \neq m}} \lambda_\ell(\textbf{q})\lambda_m(\textbf{q}), \label{eq:laginv_formulae_2}\\
s_3(\textbf{q}) & = & - \textstyle{\prod\limits_\ell} \lambda_\ell(\textbf{q}) ,\label{eq:laginv_formulae_3}
\end{eqnarray}
from which one can derive the explicit formulation in terms of the coefficients of $\mathscr{R}$ (summation over repeated indices is implied):
\begin{eqnarray}
s_1 & = & - \mathrm{tr}(\mathscr{R}) = -\theta_{\ell\ell}, \\
s_2 & = & \frac{1}{2}\left[ s_1^2 - \mathrm{tr}(\mathscr{R}^2) \right] = \frac{1}{2} \left[ s_1^2 - \theta_{\ell m}\theta_{m \ell} - \omega_{\ell m}\omega_{m \ell} \right] \nonumber\\
s_3 & = &  -\mathrm{det}(\mathscr{R}) = \frac{1}{3} \left[ -s_1^3 + 3s_1s_2 - \mathrm{tr}(\mathscr{R}^3) \right] \nonumber\\
& = & \frac{1}{3} \left[ -s_1^3 + 3s_1s_2 -\theta_{\ell m}\theta_{mn}\theta_{n\ell} -3\omega_{\ell m}\omega_{mn}\theta_{n\ell} \right]. \nonumber
\end{eqnarray}
In the previous equations, we have used the decomposition of $\mathscr{R}$ into a symmetric part $\uptheta = \left[\theta_{\ell m}\right]$ (the rate of deformation tensor) and an anti-symmetric part $\upomega = \left[ \omega_{\ell m} \right]$ (the spin tensor),
\begin{equation}
\mathscr{R}_{\ell m} = \theta_{\ell m} +\omega_{\ell m},
\end{equation}
with
\begin{eqnarray}
\theta_{\ell m} & \equiv & \frac{1}{2}(\mathscr{R}_{\ell m} + \mathscr{R}_{m \ell}) = \frac{\partial^2 \Phi_\Psi}{\partial \textbf{q}_\ell \, \partial \textbf{q}_m} , \\
\omega_{\ell m} & \equiv & \frac{1}{2}(\mathscr{R}_{\ell m} - \mathscr{R}_{m \ell}) = \frac{\partial \left( \nabla_\textbf{q} \times \textbf{A}_\Psi \right)_\ell}{\partial \textbf{q}_m}.
\end{eqnarray}
Physically, $\theta_{\ell m}$ denotes the rate of deformation of a Lagrangian fluid element and $\omega_{\ell m}$ the rate of rotation.

The surface in Lagrangian invariants space that divides real and complex solutions for $\lambda_i$ is given by the equation
\begin{equation}
27s_3^2+(4s_1^3-18s_1s_2)s_3+(4s_2^3-s_1^2s_2^2)=0.
\label{eq:discriminant_cubic}
\end{equation}
Using the discriminant of this quadratic equation, one finds two real solutions $s_{3a}$ and $s_{3b}$ ($s_{3a} \leq s_{3b}$) if and only if $s_2 \leq s_1^2/3$. Therefore, at fixed $s_1$, the regions where all the eigenvalues of $\mathscr{R}$ are real correspond to
\begin{equation}
s_2 \leq s_1^2/3, \quad s_{3a} \leq s_3 \leq s_{3b} .
\label{eq:real_solution_condition}
\end{equation}
In this case, the displacement field is purely potential and we denote the eigenvalues by $\lambda_1(\textbf{q}) \leq \lambda_2(\textbf{q}) \leq \lambda_3(\textbf{q})$. We also define the ellipticity $\varepsilon$ and prolateness $p$ \citep[e.g.][]{Peacock1985,Bardeen1986} as
\begin{eqnarray}
\varepsilon(\textbf{q}) & \equiv & \frac{\lambda_3(\textbf{q})-\lambda_1(\textbf{q})}{2\left[\lambda_1(\textbf{q})+\lambda_2(\textbf{q})+\lambda_3(\textbf{q})\right]}, \quad \mathrm{and} \\
p(\textbf{q}) & \equiv & \frac{\lambda_3(\textbf{q})-2\lambda_2(\textbf{q})+\lambda_1(\textbf{q})}{2\left[\lambda_1(\textbf{q})+\lambda_2(\textbf{q})+\lambda_3(\textbf{q})\right]}.
\end{eqnarray}
Ellipsoids are prolate-like if $-\varepsilon \leq p \leq 0$ or oblate-like if $0 \leq p \leq \varepsilon$. The limiting cases are $p=-\varepsilon$ for prolate spheroids and $p = \varepsilon$ for oblate spheroids. In the following, we will also consider $\varepsilon'(\textbf{q}) \equiv \lambda_3(\textbf{q})-\lambda_1(\textbf{q})$ and $p'(\textbf{q})~\equiv~\lambda_3(\textbf{q})-2\lambda_2(\textbf{q})+\lambda_1(\textbf{q})$, for which the boundary conditions are still valid.

After orbit-crossing, an anti-symmetric part of $\Psi$ is generated and the eigenvalues of $\mathscr{R}$ are no longer all real. We parametrize them as $\lambda_{1,2}(\textbf{q}) = a(\textbf{q}) \pm \mathrm{i}\,b(\textbf{q})$ and $\lambda_3(\textbf{q}) = c(\textbf{q})$ where $a,b,c$ are real. The invariants then simply read (see equations \eqref{eq:laginv_formulae_1}--\eqref{eq:laginv_formulae_3})
\begin{eqnarray}
s_1(\textbf{q}) & = & -2a(\textbf{q})-c(\textbf{q}) \\
s_2(\textbf{q}) & = & a^2(\textbf{q}) + b^2(\textbf{q}) + 2a(\textbf{q})c(\textbf{q}) \\
s_3(\textbf{q}) & = & -c(\textbf{q})a^2(\textbf{q})-c(\textbf{q})b^2(\textbf{q}).
\end{eqnarray}
As we will show in section \ref{sec:Lagrangian web-type classifications}, the Lagrangian invariants allow a complete classification of the cosmic web, including both potential and vortical flow patterns \citep[as with the rotational invariants, see][]{Chong1990,Wang2014}.

Finally, we note that it is possible to translate the Lagrangian invariants $s_i$ to the eigenvalues $\lambda_i$, using Vieta's formulas for cubic equations:
\begin{eqnarray}
u(\textbf{q}) & = & s_2(\textbf{q})-s_1^2(\textbf{q})/3, \\
v(\textbf{q}) & = & \frac{2}{27}s_1^3(\textbf{q})-\frac{1}{3}s_1(\textbf{q})s_2(\textbf{q})+s_3(\textbf{q}), \\
\varphi(\textbf{q}) & = & \arccos\left[ \frac{v(\textbf{q})}{2} \left( \frac{-3}{u(\textbf{q})} \right)^{3/2} \right] ;
\end{eqnarray}
and for $k \in \{0,1,2\}$,
\begin{equation}
\lambda_{k+1}(\textbf{q}) = 2 \left( \frac{-u(\textbf{q})}{3} \right)^{1/2} \cos\left( \frac{\varphi(\textbf{q})}{3} - \frac{2\pi k}{3} \right) - \frac{s_1(\textbf{q})}{3} .
\end{equation}

\subsection{Lagrangian web-type classifications}
\label{sec:Lagrangian web-type classifications}

In our earlier analysis of the volume covered by the SDSS main sample galaxies \citepalias{Leclercq2015ST}, we adopted the Eulerian cosmic web classifier known as the {\tweb} \citep{Hahn2007a,Forero-Romero2009}. In this framework, structures are classified according to the sign of the eigenvalues $\mu_1(\textbf{x}) \le \mu_2(\textbf{x}) \leq \mu_3(\textbf{x})$ of the tidal field tensor $\mathscr{T}$, defined as the Hessian of the rescaled gravitational potential $\Phi$:
\begin{equation}
\mathscr{T}_{ij} \equiv \mathrm{H}(\Phi)_{ij} = \frac{\partial^2 \Phi}{\partial \textbf{x}_i \partial \textbf{x}_j},
\end{equation}
where $\Phi$ obeys the reduced Poisson equation
\begin{equation}
\Delta \Phi(\textbf{x}) = \delta(\textbf{x}),
\end{equation}
$\delta$ being the local density contrast. A void point corresponds to no positive eigenvalue, a sheet to one, a filament to two, and a cluster to three positive eigenvalues.

Related Hessian-based techniques are the \textsc{V-web} \citep{Hoffman2012} and \textsc{classic} \citep{KitauraAngulo2012}. More broadly, Eulerian cosmic web algorithms include topological \citep{Aragon-Calvo2010a,Sousbie2011a}, stochastic \citep{GonzalezPadilla2010,Tempel2014}, and multiscale field \citep{Aragon-Calvo2007,Cautun2013,Aragon-Calvo2014} methods, all of which yield different physical insights from the phase-space analysis subject of this paper.

In this section, we discuss three alternative methods that can be used for producing \emph{Lagrangian} classifications of the cosmic web: {\diva}, {\lich} and {\origami}.

\subsubsection{DIVA}
\label{sec:DIVA}

In analogy with the \citet{Zeldovich1970} ``pancake'' theory, \citet{Lavaux2010} propose to define cosmic structures using $\lambda_1(\textbf{q}) \leq \lambda_2(\textbf{q}) \leq \lambda_3(\textbf{q})$, the eigenvalues of the shear of the displacement field, $\mathscr{R}$. In the {\diva} (DynamIcal Void Analysis) scheme, a particle is defined as belonging to a Lagrangian void, sheet, filament or cluster, if, respectively, zero, one, two or three of the $\lambda_i$ are negative. In order to deal only with real eigenvalues, {\diva} assumes that the flow is potential, i.e. that $\mathscr{R}$ is equal to its symmetric part $\uptheta$.

The {\diva} and {\tweb} methods share strong similarity in their formalism. However, a major difference is that the {\tweb} operates on \textit{voxels} of the discretized domain, whereas {\diva} operates on \textit{Lagrangian patches} (numerically approximated by particles). As shown in \citetalias{Leclercq2015ST} (sections III and IV), the {\tweb} can be used to classify both early-time and late-time structures; however the {\tweb} classification of early-time structures is performed at the level of voxels of the discretized initial density field, whereas the {\diva} classification of Lagrangian structures is performed at the level of particles of the dark matter sheet.

\subsubsection{LICH}
\label{sec:LICH}

\begin{figure*}
\begin{center}
\includegraphics[width=\textwidth]{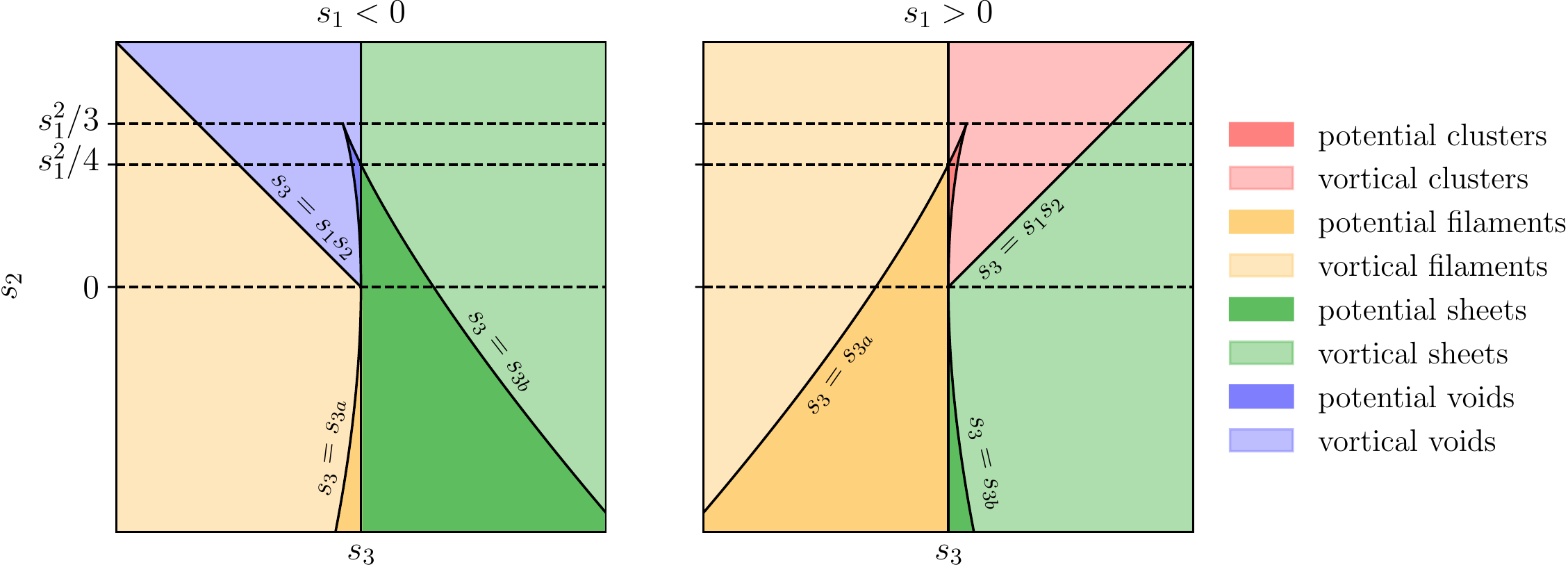}
\caption{Illustration of structure type classification in Lagrangian invariants space, according to the {\lich} procedure \citep[see also figure 3 in][]{Wang2014}. We show the $s_3-s_2$ plane for $s_1<0$ (left panel) and $s_1>0$ (right panel). The Lagrangian structure type corresponding to each type of potential or vortical flow is given in the legend.\label{fig:laginv}}
\end{center}
\end{figure*}

This section introduces a new cosmic-web classification procedure based on the Lagrangian invariants, which we call {\lich} (Lagrangian Invariants Classification of Heterogeneous flows). As we will show, {\lich} is a generalization of {\diva} to the case of heteregenous (i.e. potential and vortical) flows. {\lich} uses for the Lagrangian invariants the same formalism as introduced by \cite{Wang2014} for the rotational invariants; the reader is therefore referred to this paper for more details.

Let us first consider the case of a potential flow. The real solution condition, equation \eqref{eq:real_solution_condition}, must be fulfilled. The case where all the $\lambda_i$ are negative (a {\diva} cluster) further corresponds to $s_i>0$ for all $i \in \{1,2,3\}$ (see equations \eqref{eq:laginv_formulae_1}--\eqref{eq:laginv_formulae_3}). A {\lich} \textit{potential cluster} is therefore defined by
\begin{equation}
s_1\geq 0, \quad 0 \leq s_2 \leq s_1^2/3, \quad \mathrm{max}(s_{3a},0) \leq s_3 \leq s_{3b} .
\end{equation}
Similarly, a {\lich} \textit{potential void}, which generalizes a {\diva} void (all $\lambda_i$ positive), corresponds to 
\begin{equation}
s_1 \leq 0, \quad 0 \leq s_2 \leq s_1^2/3, \quad s_{3a} \leq s_3 \leq \mathrm{min}(s_{3b},0) .
\end{equation}
Two regions in Lagrangian invariants space correspond to a {\diva} sheet ($\lambda_1, \lambda_2 \leq 0$, $\lambda_3 \geq 0$):
\begin{eqnarray}
s_1 & \leq & 0, \quad s_2 \leq s_1^2/4, \quad 0 \leq s_3 \leq s_{3b}; \\
s_1 & \geq & 0, \quad s_2 \leq 0, \quad 0 \leq s_3 \leq s_{3b} . 
\end{eqnarray}
They define the {\lich} \textit{potential sheets}. Two regions also correspond to a {\diva} filament ($\lambda_1 \leq 0$, $\lambda_2, \lambda_3 \geq 0$):
\begin{eqnarray}
s_1 & \leq & 0, \quad s_2 \leq 0, \quad s_{3a} \leq s_3 \leq 0;\\
s_1 & \geq & 0, \quad s_2 \leq s_1^2/4, \quad s_{3a} \leq s_3 \leq 0.
\end{eqnarray}
They define the {\lich} \textit{potential filaments}.

The case of a vortical flow corresponds to either $s_2~>~s_1^2/3$ (in which case there is no solution to equation \eqref{eq:discriminant_cubic}), either $s_3~\not\in~[s_{3a}, s_{3b}]$. In these regions in Lagrangian invariants space, an additional frontier is important: $\lambda_1=-\lambda_2$ (i.e. $a=0$), which corresponds to the plane
\begin{equation}
s_3 = s_1 s_2.
\end{equation}
The remaining regions can then be labeled as \textit{vortical voids}, \textit{vortical sheets}, \textit{vortical filaments} or \textit{vortical clusters}. Figure \ref{fig:laginv} summarizes the {\lich} classification procedure in Lagrangian invariants space. For the detailed criteria defining all the regions there, the reader is referred to Appendix A in \cite{Wang2014}.

In section \ref{sec:Lagrangian classifications}, we will also use a simplified version of the full {\lich} procedure, where particles fall into one of the usual four categories (void, sheet, filament or cluster) by simply ignoring the potential/vortical subdivision. Note that this simplified {\lich} procedure is different from {\diva}, since it does not use the initial assumption that the flow is potential.

\subsubsection{ORIGAMI}
\label{sec:ORIGAMI}

Another way of classifying particles is to consider them as vertices of an initially regular grid, distorted by gravity as the large-scale structure forms. Shell-crossing happens when Lagrangian cells collapse or invert. A particle's {\origami} (Order-ReversIng Gravity, Apprehended Mangling Indices) morphology is defined by the number of orthogonal axes along which shell-crossing has occurred \citep{KnebeEtAl2011,Falck2012}. More precisely, void, sheet, filament, and cluster particles are defined as particles that have been crossed along zero, one, two, and three orthogonal axes, respectively.

In practice, the computer implementation of {\origami} works as follows: a particle $\ell$ has been crossed along an axis $\widehat{\textbf{n}}$ if there exists a particle $m$ in the same row of the Lagrangian grid, such that $(\textbf{q}_\ell-\textbf{q}_m) \cdot \widehat{\textbf{n}}$ and $(\textbf{x}_\ell-\textbf{x}_m) \cdot \widehat{\textbf{n}}$ have opposite sign. Such particle-crossing are checked along the three orthogonal axes of the original Cartesian grid, as well as three triplets of orthogonal axes, obtained by rotating the original grid by 45$\degree$ along one of its axes. For more details, the reader is referred to section 2.1 in \citet{Falck2012}.

\subsection{Tessellation of the Lagrangian grid}
\label{sec:Tessellation of the Lagrangian grid}

\begin{figure}
\begin{center}
\includegraphics[width=0.7\columnwidth]{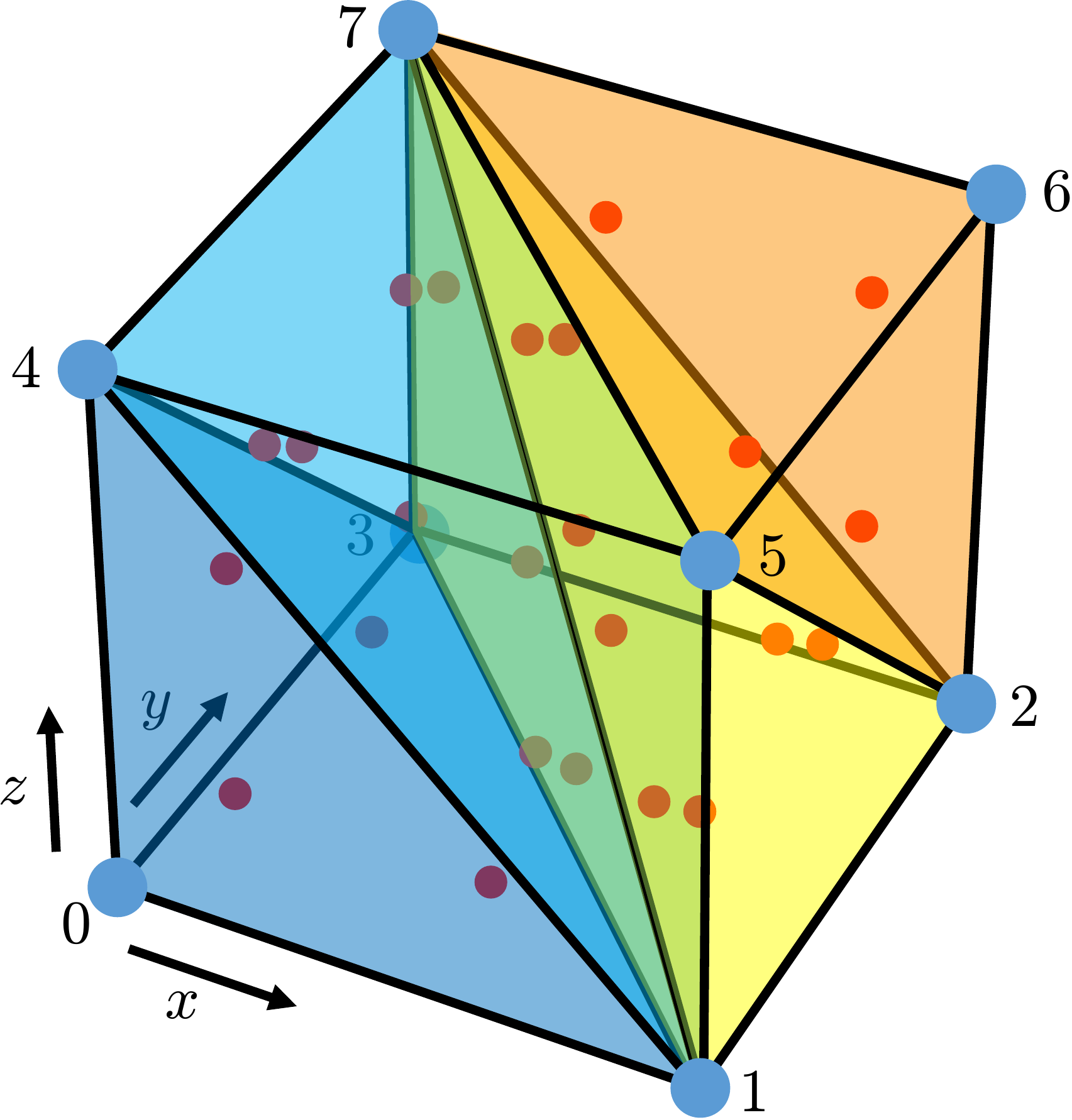}
\caption{Tessellation of the unit cube into six tetrahedra. The blue points represent the usual dark matters particles (the \textit{flow tracers}), initially placed at the vertices of a uniform Lagrangian lattice. Each of the six tetrahedra can be approximated by four pseudo-particles shown in red (the \textit{mass tracers}), which carry 1/24 of the mass of one flow tracer.\label{fig:illustration_tessellation}}
\end{center}
\end{figure}

\cite{Abel2012} and \cite{Shandarin2012} proposed new techniques to trace the dark matter phase-space sheet and demonstrated that they give a wealth of information previously unavailable in simulations. In this paper, we will demonstrate that these techniques can now be used in data-constrained realizations of the nearby Universe. The core idea is to use a tessellation of the uniform Lagrangian lattice of particles. Each fundamental cube is decomposed into tetrahedral cells that use the particles as their vertices. These cells are tracked in the final conditions, using the particle IDs which are set-up so that they encode the initial position of the particle.

Several possible decompositions of the unit cube exist: for example, \cite{Shandarin2012} use five tetrahedra of two different volumes. In this work, however, we will follow \cite{Abel2012,HahnAnguloAbel2015} and use the Delaunay decomposition into six identical tetrahedra (see figure \ref{fig:illustration_tessellation}). Specifically, the connectivity list is $[0,1,3,4]$, $[1,3,4,7]$, $[1,4,5,7]$, $[1,2,5,7]$, $[1,2,3,7]$ and $[2,5,6,7]$ where the unit cube vertices are labeled as in figure \ref{fig:illustration_tessellation}.

Initially, the volume of each tetrahedron is $d^3/6$ where $d$ is the side length of the fundamental cube. In the final state, the same tetrahedron has a volume given by a determinant (or cross-product)
\begin{equation}
V_\mathrm{tet} \equiv \frac{\det\left| \textbf{AB},\textbf{AC},\textbf{AD} \right|}{6} = \frac{\textbf{AB} \cdot (\textbf{AC} \times \textbf{AD})}{6} ,
\end{equation}
where A, B, C, D label the final positions of the tetrahedron vertices. Note that the volume can be negative if the tetrahedron has been turned inside-out. In fact, the number of sign reversals of an elementary Lagrangian volume defines the flip-flop field \citep[e.g.][]{Neyrinck2012,ShandarinMedvedev2016}. Each tetrahedron carries 1/6 of the mass of a particle $m_\mathrm{p}$, the mass density associated is therefore $\rho_\mathrm{tet} = m_\mathrm{p}/\!\left(6 |V_\mathrm{tet}|\right)$. At any particle position, there are 24 tetrahedra that share this vertex ($2 \times 4 + 4 \times 4$, see figure \ref{fig:illustration_tessellation}). 

Tessellating the dark matter phase-space sheet allows one to define a new quantity on the Lagrangian grid, the \textit{primordial stream density},
\begin{equation}
\mathrm{psd}(\textbf{q}) \equiv \frac{4 \, d^3}{\sum_{i=1}^{24} |V_\mathrm{tet}^i(\textbf{q})|}.
\label{eq:psd_def}
\end{equation}
where the summation runs on the 24 tetrahedra that share the particle's initial position $\textbf{q}$ as one of their vertices. Altogether, these tetrahedra have an initial volume of $24 \times d^3/6 = 4\, d^3$.

In addition to the primordial stream density, tracking the final configuration of the tessellated dark matter phase-space sheet allows one to perform the Lagrangian transport of any quantity defined on a particle basis, which can be used to define a variety of estimators on the Eulerian grid. This is achieved by finding all the tetrahedra that contain the point $\textbf{x}$ at which properties are to be determined. Each such tetrahedron is given four coefficients $f_\mathrm{A}$, $f_\mathrm{B}$, $f_\mathrm{C}$, $f_\mathrm{D}$ for the four particles that constitute its vertices (A, B, C, D). Fields are interpolated inside each tetrahedron using Shepard's method (inverse distance-squared weighting); namely, each tetrahedron deposits to the point $\textbf{x}$ the quantity
\begin{equation}
\sum_{\mathrm{P} \in \{\mathrm{A}, \mathrm{B}, \mathrm{C}, \mathrm{D}\} } \frac{w_\mathrm{P} \times f_\mathrm{P}}{w},
\end{equation}
where $w_\mathrm{P} \equiv 1/||\textbf{x}-\textbf{x}_\mathrm{P}||^2$ and $w \equiv \sum_\mathrm{P} w_\mathrm{P}$.

In this work, we consider the following coefficients and associated estimators:
\begin{itemize}
\item $f_\mathrm{A} = f_\mathrm{B} = f_\mathrm{C} = f_\mathrm{D} = 1$ simply yields the number of tetrahedra enclosing point $\textbf{x}$. Since we start from a complete tessellation of the considered volume, this is also the number of matter streams contributing at point $\textbf{x}$, which we call the Eulerian \textit{secondary stream density}, following \citet{Abel2012}.
\item $f_\mathrm{A} = f_\mathrm{B} = f_\mathrm{C} = f_\mathrm{D} = \rho_\mathrm{tet}$ yields an estimator of the final Eulerian \textit{density field} \citep[see][]{Abel2012}, which we note $\widehat{\rho}_\mathrm{tet}$. 
\item $f_\mathrm{P} = \rho_\mathrm{tet}$ if particle P is of type $\mathrm{T}_i$, 0 otherwise yields an estimator of the density contained in structure of type $\mathrm{T}_i$, i.e. $\widehat{\rho}_\mathrm{tet}(\mathrm{T}=\mathrm{T}_i)$. The field $\widehat{\rho}_\mathrm{tet}(\mathrm{T}=\mathrm{T}_i)/\widehat{\rho}_\mathrm{tet}$ represents the probability for a final Eulerian point to belong to structure type $\mathrm{T}_i$. This method can therefore be used to \textit{translate Lagrangian classifications into Eulerian coordinates}. As it is computationally expensive, we will introduce a simplified alternative in section \ref{sec:Translation of Lagrangian classifications to Eulerian coordinates}.
\item $f_\mathrm{P} = v_s(\mathrm{P}) \times \rho_\mathrm{tet}$ where $v_s(\mathrm{P})$ is the velocity of particle P along the direction $s \in \{ x, y, z\}$ yields an estimator of the component $s$ of the mass-weighted velocity field. The resulting field divided by $\widehat{\rho}_\mathrm{tet}$ yields the \textit{velocity field} estimated from the dark matter phase-space sheet \citep[see][]{HahnAnguloAbel2015}.
\end{itemize}

Instead of explicitly projecting all the tetrahedra to the Eulerian grid, it is possible to approximate each of them by four pseudo-particles \citep[in red in figure \ref{fig:illustration_tessellation}; see the discussion in][section 2.4, for where to place these pseudo-particles]{Hahn2013}. In this context, the original particles are called the \textit{flow tracers} and the pseudo-particles are called the \textit{mass tracers}. Flow tracers uniquely determine the position of mass tracers, which can be deposited to the Eulerian grid by standard interpolation techniques such as the cloud-in-cell \citep[CiC,][]{Hockney1981} algorithm.

\section{Lagrangian analysis of the SDSS volume}
\label{sec:Lagrangian analysis of the SDSS volume}

As mentioned in the introduction, this paper is part of a series following the recent application of the Bayesian inference framework {\borg} to the SDSS main sample galaxies \citepalias{Jasche2015BORGSDSS}. We further rely on the set of 1,097 constrained simulations presented in \citetalias{Leclercq2015ST}. These realizations are obtained, starting from {\borg} initial conditions, by non-linear filtering with {\cola} \citep{Tassev2013}. The initial density field, defined on a $256^3$-voxel grid with side length of $750$~Mpc/$h$, is populated by $512^3$ dark matter particles placed on a regular Lagrangian lattice. Particles are evolved with 2LPT to the redshift of $z=69$ then with 30 {\cola} timesteps from $z=69$ to $z=0$. The displacement field is obtained on the Lagrangian lattice by subtracting the final position of particles from their initial position, and any necessary quantity involving derivatives is computed using FFTs on the same lattice.

In this section, we first analyze the Lagrangian properties of individual samples (section \ref{sec:Analysis of individual samples}), then show how the full set of constrained samples can be used to propagate uncertainties to all inferred quantities (section \ref{sec:Uncertainty quantification}).

\subsection{Analysis of individual samples}
\label{sec:Analysis of individual samples}

\begin{figure*}
\begin{center}
\includegraphics[width=\textwidth]{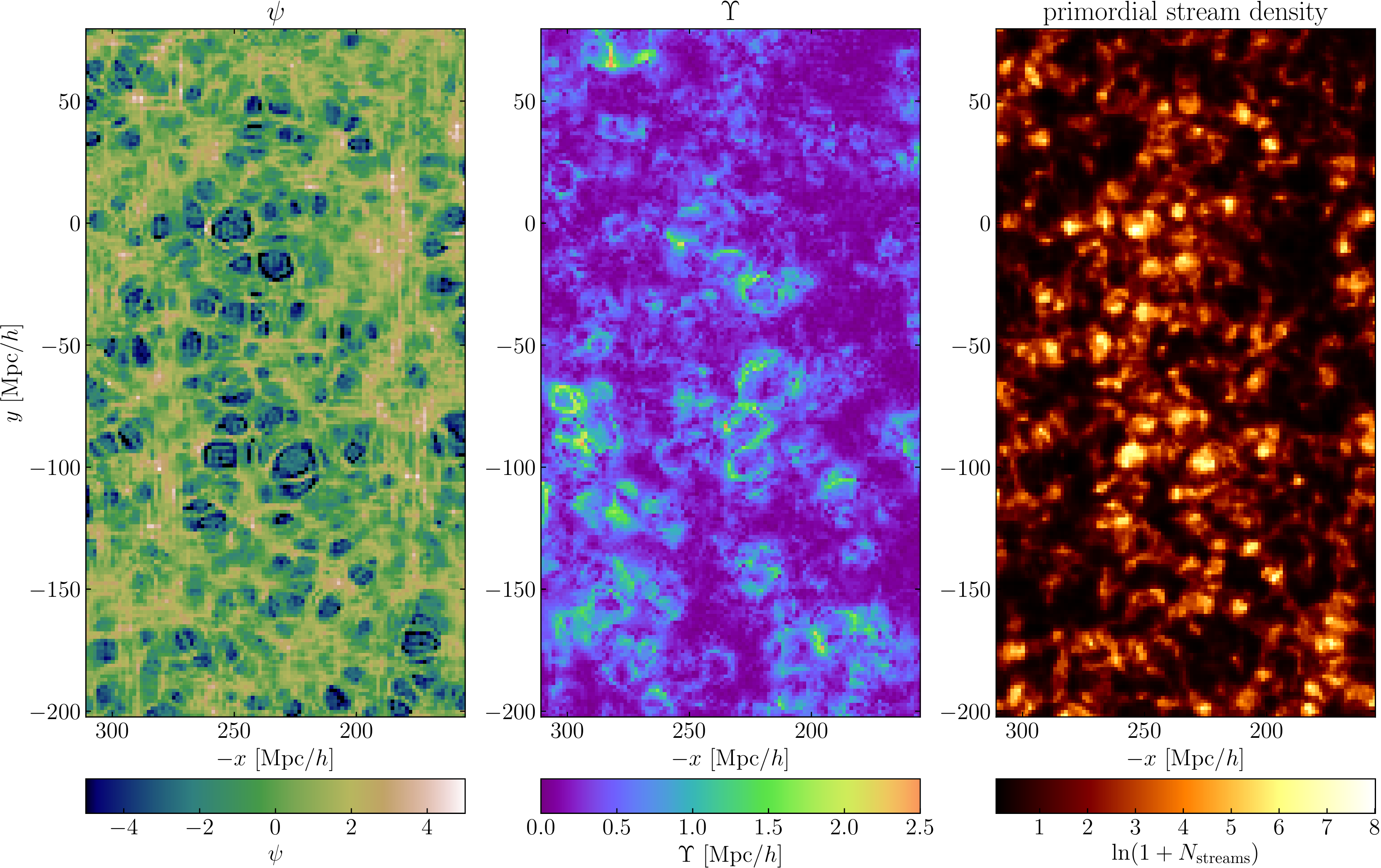}\vspace{4pt}
\includegraphics[width=\textwidth]{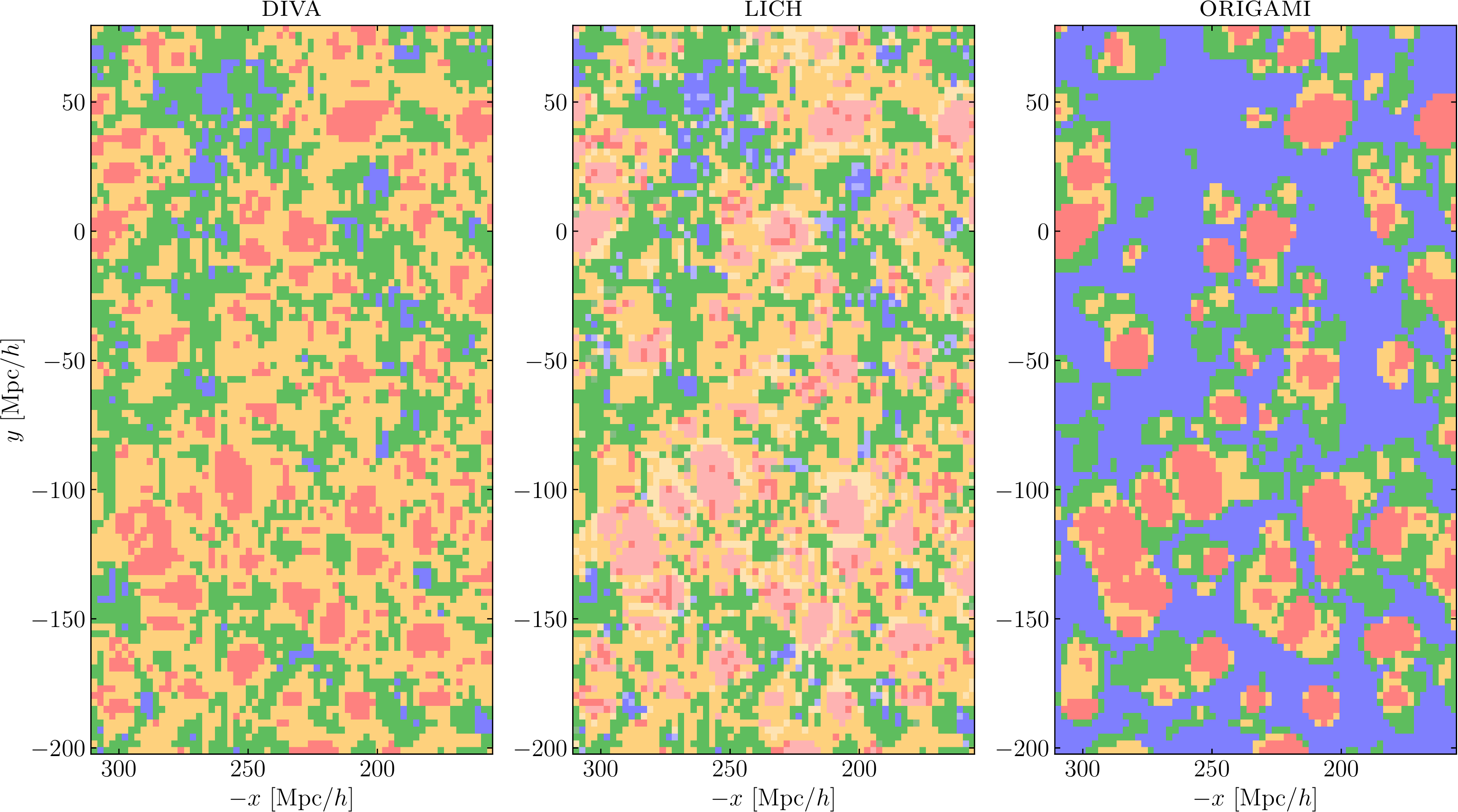}
\caption{Slices through one sample of the observed nearby Lagrangian dark matter sheet, as constrained by the SDSS main galaxy sample. A region of the celestial equatorial plane encompassing the Sloan Great Wall in shown. Each pixel corresponds to one particle of the Lagrangian lattice. The first row shows the divergence of the displacement field $\psi$ (left panel), the norm of the vector part of the displacement field $\Upsilon$ (middle panel), and the primordial stream density, defined by equation \eqref{eq:psd_def} (right panel). The second row shows Lagrangian classifications of particles: {\diva} (left panel), {\lich} (middle panel) and {\origami} (right panel). Blue, green, yellow, and red correspond to void, sheet, filament, and cluster particles, respectively. For {\lich}, dark colors correspond to potential structures and light colors to vortical structures, as in figure \ref{fig:laginv}.\label{fig:dmsheet_one_sample}}
\end{center}
\end{figure*}

\subsubsection{Stream properties}
\label{sec:Stream properties}

The first row of figure \ref{fig:dmsheet_one_sample} shows the stream properties in one of the samples of our set. The slices show a region of the celestial equatorial plane encompassing the Sloan Great Wall \citep{Gott2005,Einasto2011}.\footnote{In the plots of this paper, we kept the coordinate system of \citetalias{Jasche2015BORGSDSS}.} The flow is characterized in terms of the divergence of the displacement field $\psi(\textbf{q}) \equiv~\nabla_\textbf{q} \cdot \Psi(\textbf{q})$, the norm of its vector part $\Upsilon(\textbf{q}) \equiv \lVert \nabla_\textbf{q} \times \textbf{A}_\Psi \rVert$, and the primordial stream density $\mathrm{psd}(\textbf{q})$ (see equation \eqref{eq:psd_def}). $\psi$ quantifies the angle-average stretching and distortion of the dark matter sheet \citep{Neyrinck2013} and contains all the information on $\Psi$ as long as the flow remains potential. $\Upsilon$ can be used to quantify how much vorticity is generated when structure formation enters the multi-stream regime, and the primordial stream density counts these streams. 

It is important to note that, since the displacement field is curl-free up to second order in Lagrangian perturbation theory \citep[see e.g.][]{Bernardeau2002}, and since {\borg} relies on 2LPT as a proxy for gravitational dynamics \citep{Jasche2013BORG}, any vorticity present in our large-scale structure realizations is not explicitly constrained by the data. The $\Upsilon$ field therefore corresponds one plausible realization, compatible both with the constrained initial conditions and with the non-linear structure formation model ({\cola}).

Our constrained samples exhibit the same behavior as previously observed in simulations. Specifically, regions of high primordial stream density nicely correspond to Lagrangian clusters, the ``lakes'' where $\psi \approx -3$. This is also where a rotational part of the displacement field is generated. Conversely, single-stream regions correspond to the ``mountains'' where $\psi>0$. From the behavior of the $\Upsilon$ field (limited at most to $3$~Mpc/$h$ in the regions of highest stream density), we also note that we can take advantage of the absence of substantial vorticity on large scales. This is consistent with the results of \citet{Chan2014}: at $z=0$ and up to $k \approx 1$~Mpc/$h$, the contribution to the power spectrum of $\Psi$ coming from its curl component remains negligible compared to the non-linear evolution of its scalar part. Therefore, for the sake of our analysis (except when explicitly measuring vorticity), it will be safe to assume that all the information on $\Psi$ is contained in its divergence $\psi$, or equivalently, that $\mathscr{R}=\uptheta$.

\subsubsection{Lagrangian classifications}
\label{sec:Lagrangian classifications}

We then assign a Lagrangian structure type to each particle, according to the {\diva}, {\lich}, and {\origami} procedures. \citet{Lavaux2010} discuss the possibility of smoothing the displacement field at different Lagrangian scales, in order to probe the hierarchical cosmic web. This allows the description of structure formation at different dynamical epochs, each Lagrangian scale corresponding to a different collapse time. In contrast, explicit Eulerian multiscale formalisms \citep{Aragon-Calvo2007,Cautun2013,Aragon-Calvo2010a,Aragon-Calvo2013} probe the hierarchy of structures at a given time. The adhesion approximation \citep{Gurbatov1989,Weinberg1990,Kofman1990,Kofman1992,Gurbatov2012,Hidding2016} provides a Lagrangian description of the hierarchical evolution of structures, but it is only correct in the linear and weakly non-linear regime, contrary to the numerical {\cola} model used in this work. The main effect of smoothing the Lagrangian grid is to change the unconditional number of particles in different structures; data constraints do not dramatically alter this result \citep{Lavaux2010}. Therefore, while this would be perfectly feasible in our framework, we choose to focus our analysis on the scale defined by our setup, without an additional smoothing step. This choice corresponds to a resolution of $\sim 1.5$~Mpc/$h$ for the Lagrangian grid ($512^3$ particles in a $750$~Mpc/$h$ cubic box) and $\sim 3$~Mpc/$h$ for discretized Eulerian maps (on a $256^3$-voxel grid).

The second row of figure \ref{fig:dmsheet_one_sample} represents Lagrangian structures on the grid of particles, in one of our constrained samples. From left to right, the panels show the {\diva}, {\lich}, and {\origami} morphologies of the dark matter sheet. As expected, all the classifications correlate well with the stream properties shown in the first panel (in particular with $\psi$). Clusters (in red) correspond to the ``lakes'' of high primordial stream density where $\psi \approx -3$, while less complex structures are regions with fewer streams and higher values of $\psi$. Single-stream regions, which correspond to {\origami} void particles, form a percolating system, as noted by \citet{Falck2015}.

The difference between {\diva} and {\origami} classifications (notable primarily for voids, then for sheets and filaments) is caused by the different phase-space criterion used for defining structures: particle-crossings that have already happened for {\origami}, versus those non-locally predicted from the deformation tensor for {\diva}. A large single-stream void percolates with {\origami}, while filament and sheet regions are limited: this reflects the current configuration of the cosmic web. Conversely, with {\diva}, particles around filaments and sheets typically get the signature of these structures, on the basis of the dynamical trend.

As shown in section \ref{sec:LICH}, {\lich} is a generalization of {\diva} to the case of heterogeneous (potential and vortical) flows --- applying {\lich} under the assumption of a potential flow yields {\diva}. Thus, the comparison between {\diva} and {\lich} allows to check this assumption. Comparing the {\diva}, {\lich}, and $\Upsilon$ panels shows that the two classifications differ in the regions where vorticity is generated in $\Psi$. Specifically, with {\lich}, most cluster regions are vortical ($80.7\%$ of cluster particles), except the smallest ones, and surrounded by a filamentary shell. $19.6\%$ of filament particles are vortical, in the surroundings clusters; the rest are potential. Sheet regions are potential ($96.0\%$ of particles). Void regions are mostly potential ($65.8\%$ of particles), but with some spinning regions. When ignoring the subclassification into potential and vortical structures, but without assuming a potential flow, the simplified {\lich} procedure agrees with {\diva} for $91.6\%$ of particles, which is an additional argument for the accuracy of the potential flow approximation. The difference is mostly driven by clusters and voids, with $18.8\%$ of {\lich} cluster and $16.5\%$ of {\lich} void particles misidentified by {\diva} (almost all of which as filaments and sheets, respectively).

\subsection{Uncertainty quantification}
\label{sec:Uncertainty quantification}

\begin{figure*}
\begin{center}
\includegraphics[width=\textwidth]{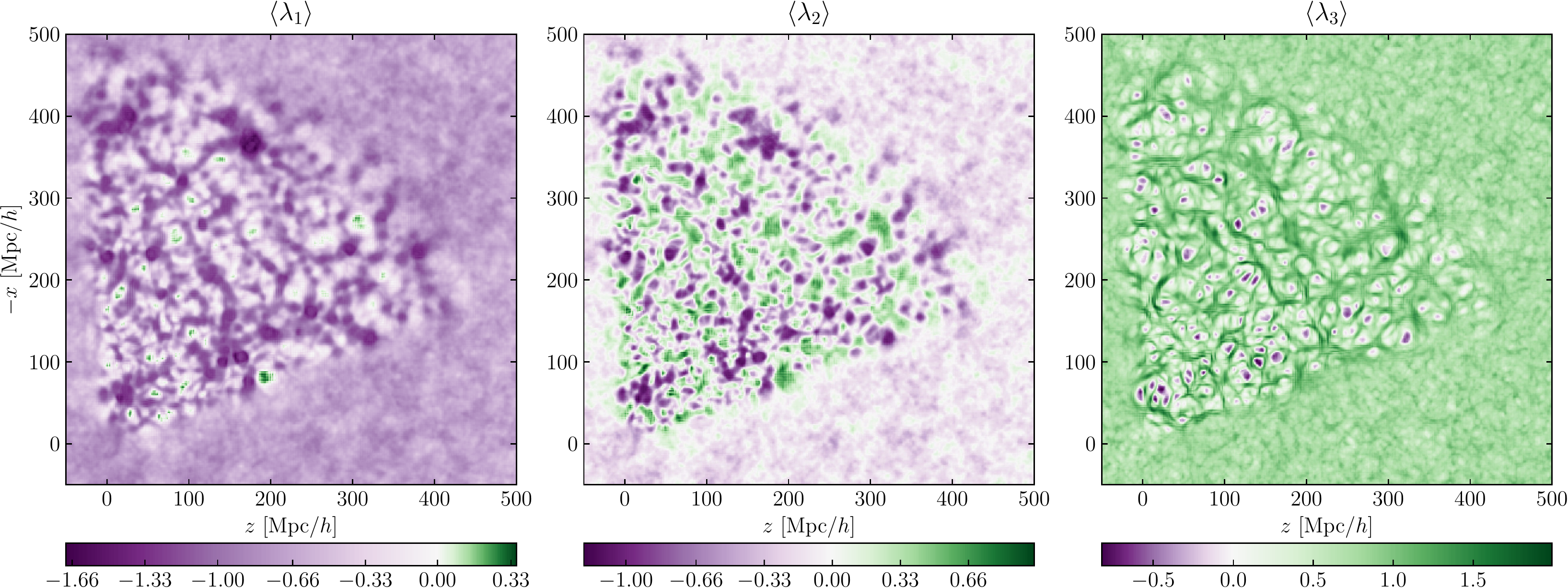}\vspace{8pt}
\includegraphics[width=\textwidth]{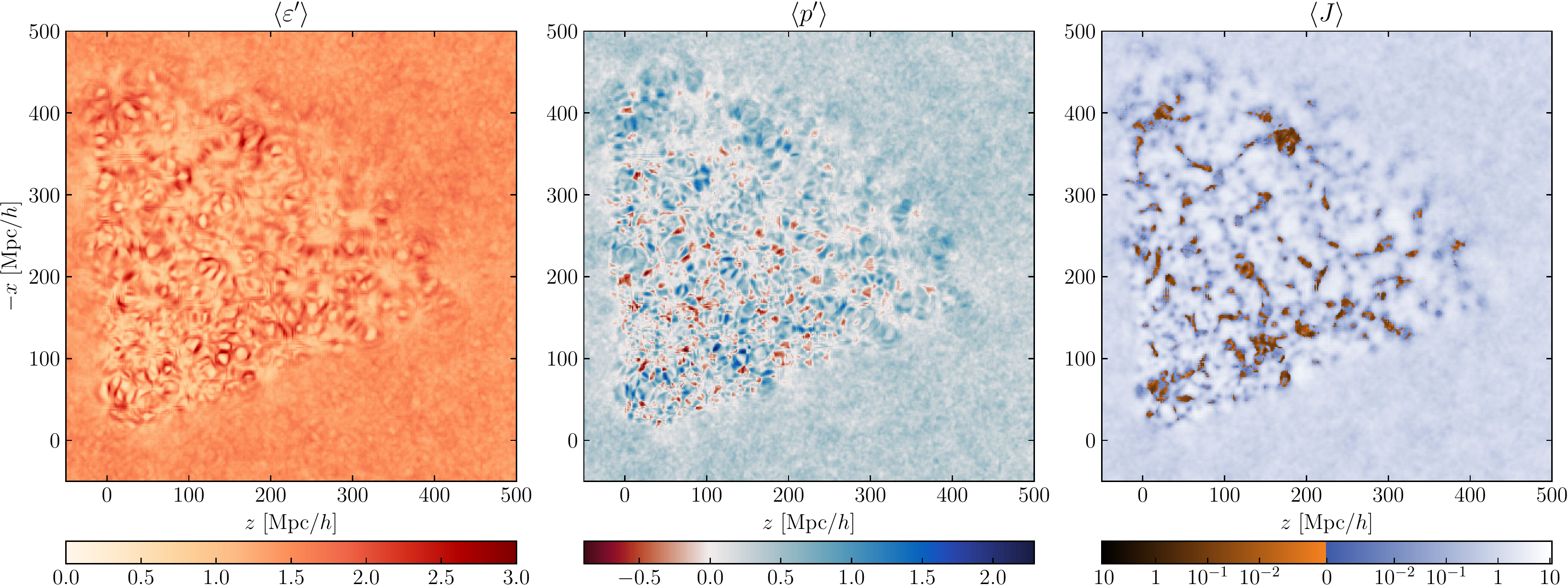}\vspace{8pt}
\includegraphics[width=\textwidth]{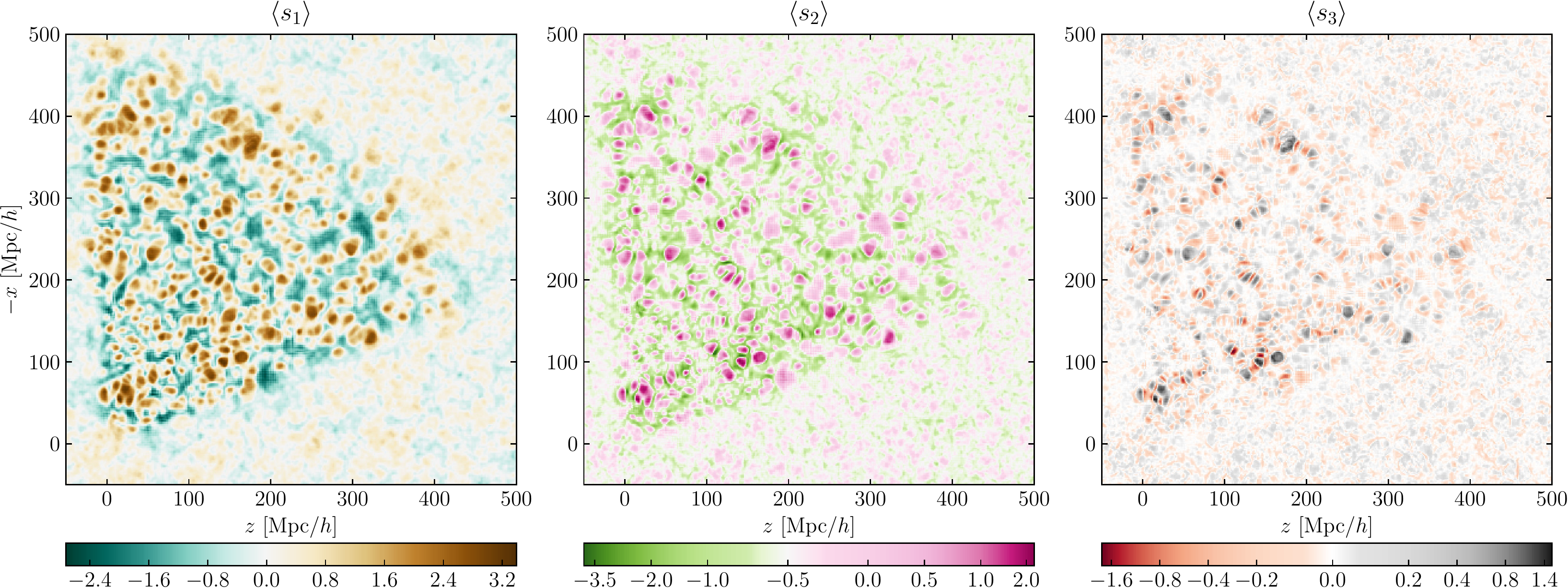}
\caption{Uncertainty quantification on the observed Lagrangian dark matter sheet. The first row shows slices through the posterior mean of the eigenvalues $\lambda_1(\textbf{q}) \leq \lambda_2(\textbf{q}) \leq \lambda_3(\textbf{q})$ of the rate of deformation tensor, used by {\diva}. The second row shows the ellipticity (left panel) and prolateness (middle panel) of the Lagrangian potential. The right panel shows the mean of the Jacobian $J(\textbf{q})$, with particles colored according to their Eulerian volume times their parity. The third row shows the posterior mean of the Lagrangian invariants, used by {\lich}.\label{fig:diva_lambda_mean}}
\end{center}
\end{figure*}

As in our previous work \citep{Jasche2015BORGSDSS,Leclercq2015ST,Lavaux2016BORG2MPP}, the variation among constrained samples constitutes a Bayesian quantification of uncertainties (coming in particular from the survey mask and selection effects). This is illustrated in figure \ref{fig:diva_lambda_mean}, where we show the particle-wise posterior mean of different quantities on the Lagrangian grid. The first row shows the eigenvalues $\lambda_1(\textbf{q}) \leq \lambda_2(\textbf{q}) \leq \lambda_3(\textbf{q})$ of the rate of deformation tensor $\uptheta$ (the symmetric part of the shear of the displacement field $\mathscr{R}$). These are used by {\diva} to classify structures. The second row shows derived quantities: the (reduced) ellipticity $\varepsilon'(\textbf{q}) \equiv \lambda_3(\textbf{q}) - \lambda_1(\textbf{q})$ and prolateness $p'(\textbf{q}) \equiv \lambda_3(\textbf{q}) - 2\lambda_2(\textbf{q}) + \lambda_1(\textbf{q})$, and the Jacobian of the transformation from Lagrangian to Eulerian coordinates, $J(\textbf{q}) = (1+\lambda_1(\textbf{q}))(1+\lambda_2(\textbf{q}))(1+\lambda_3(\textbf{q}))$. For each particle, it is the volume of the associated fluid element times its parity \citep{Neyrinck2012,ShandarinMedvedev2016}. White/blue particles have the same parity as in the initial conditions, while black/orange particles have swapped parity. Caustics are clearly visible wherever blue and orange regions are juxtaposed, for example in the patches that collapse to form halos and in the shell-crossed walls surrounding voids. In some clusters, several parity inversions can be observed. The last row shows the posterior mean for the Lagrangian invariants $s_1(\textbf{q})$, $s_2(\textbf{q})$, $s_3(\textbf{q})$, which are used by {\lich} to classify structures. These cosmographic results constitute the first characterization of the nearby Lagrangian dark matter sheet, and contain a wealth of information on the dynamic large-scale environment of SDSS main sample galaxies. They come as fully probabilistic, which means that uncertainties are quantified at the level of the map.

\begin{figure*}
\begin{center}
{\diva}
\includegraphics[width=\textwidth]{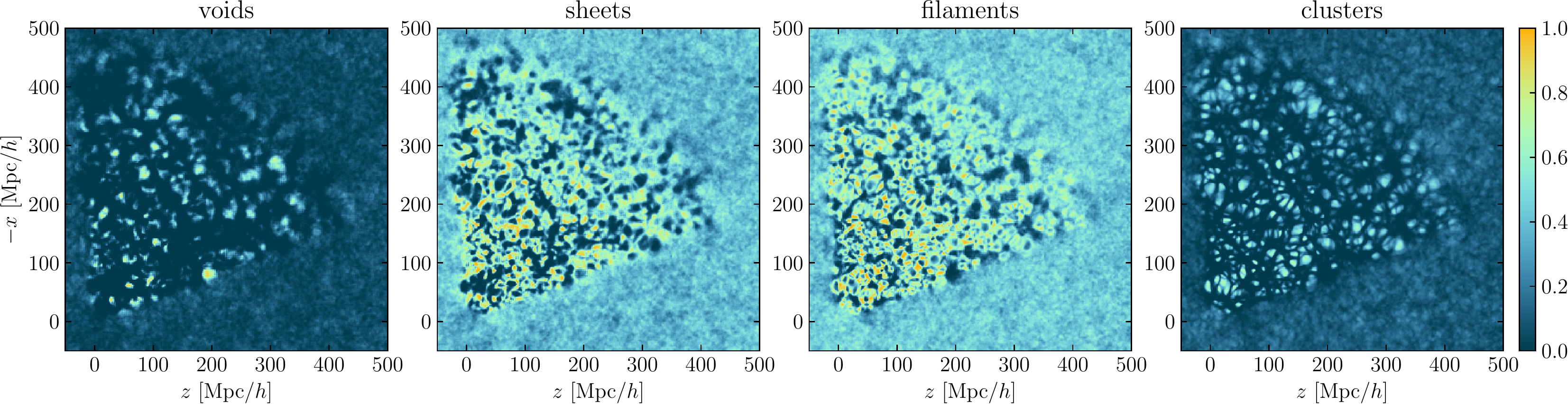}
\rule{4cm}{0.4pt}\\
{\lich}
\includegraphics[width=\textwidth]{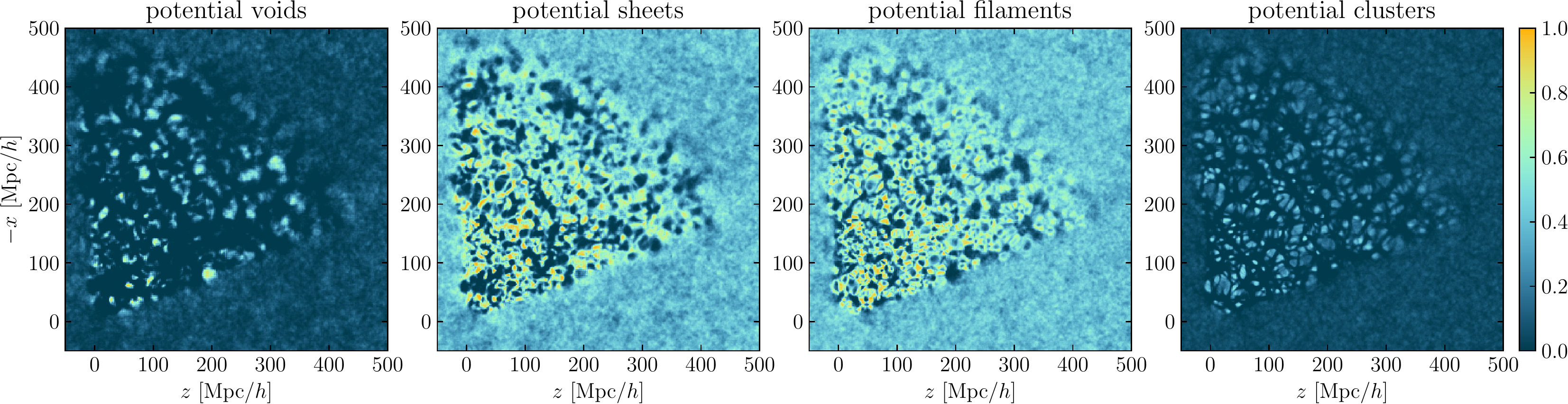}
\includegraphics[width=\textwidth]{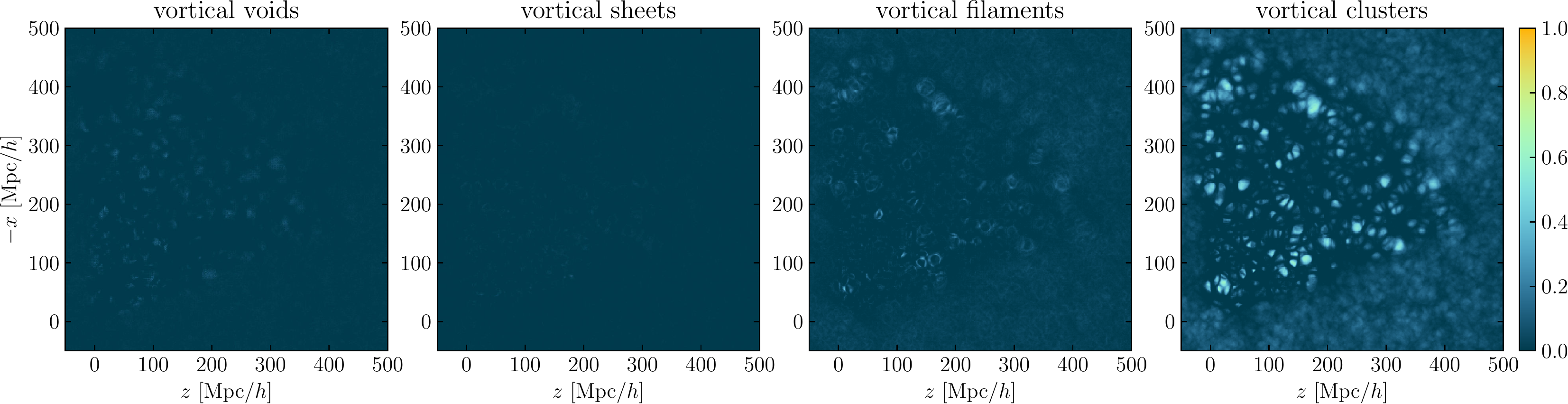}
\rule{4cm}{0.4pt}\\
{\origami}
\includegraphics[width=\textwidth]{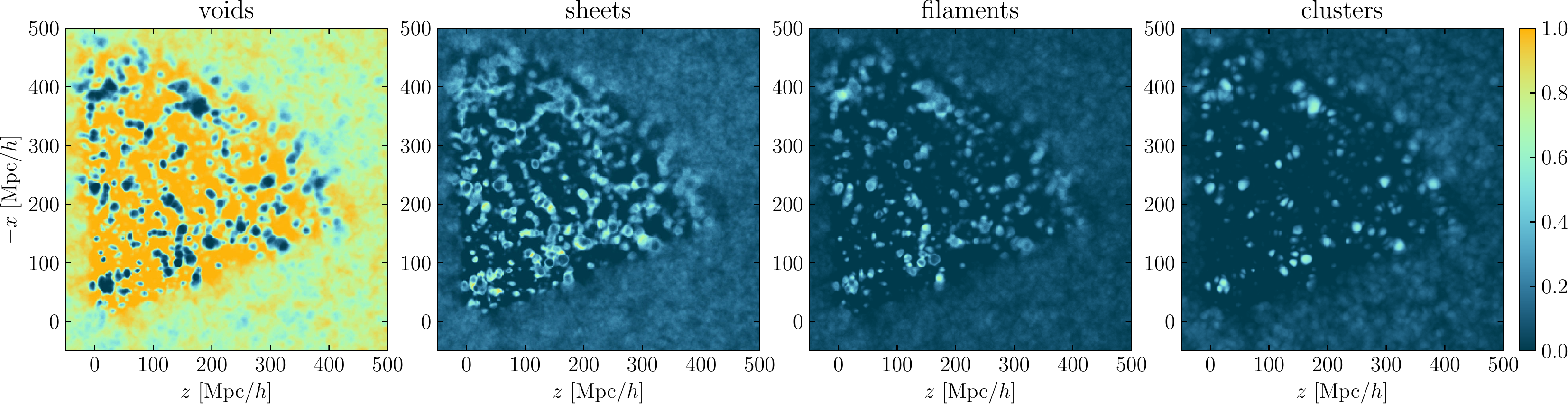}
\caption{Slices through the posterior probabilities for different structure types (from left to right: void, sheet, filament, and cluster), in the primordial large-scale structure in the Sloan volume ($a=10^{-3}$). Structures are defined on the Lagrangian grid of $512^3$ particles with {\diva} (first row), {\lich} (second and third rows), or {\origami} (fourth row). For {\lich}, potential and vortical structures are distinguished. For each classifier, the four or eight three-dimensional probabilities sum up to one at each location. For comparison, the reader is referred to figure 6 in \citet{Leclercq2015ST}, where the primordial density field, defined on a $256^3$-voxel grid, is classified according to the {\tweb} procedure.\label{fig:pdf_initial}}
\end{center}
\end{figure*}

Uncertainty quantification can also be self-consistently propagated to Lagrangian structure type classification, as we now show. Let us denote by $\xi$ one of the Lagrangian classifiers ({\diva}, {\lich}, or {\origami). In a specific realization, $\xi$ uniquely characterizes the Lagrangian grid, meaning that it provides $N_\mathrm{T}$ particle-wise scalar fields that obey the following conditions for each particle $\textbf{q}_\ell$:
\begin{equation}
\mathrm{T}_i(\textbf{q}_\ell|\xi) \in \{0,1\} \; \mathrm{for} \; i \in \llbracket 0,N_\mathrm{T}-1 \rrbracket; \; \sum\limits_{i=0}^{N_\mathrm{T}-1} \mathrm{T}_i(\textbf{q}_\ell|\xi) = 1 
\end{equation}
where $N_\mathrm{T}=4$ for {\diva} and {\origami} ($\mathrm{T}_0=$ void, $\mathrm{T}_1=$ sheet, $\mathrm{T}_2=$ filament, $\mathrm{T}_3=$ cluster); $N_\mathrm{T}=8$ for {\lich} ($\mathrm{T}_0=$ potential void, $\mathrm{T}_1=$ potential sheet, $\mathrm{T}_2=$ potential filament, $\mathrm{T}_3=$ potential cluster, $\mathrm{T}_4=$ vortical void, $\mathrm{T}_5=$ vortical sheet, $\mathrm{T}_6=$ vortical filament, $\mathrm{T}_7=$ vortical cluster). By applying $\xi$ to the complete set of {\borg}-{\cola} realizations, we can quantify the degree of belief in structure type classification in terms of $N_\mathrm{T}$ particle-wise scalar fields that obey the following conditions for each particle $\textbf{q}_\ell$:
\begin{equation}
\mathcal{T}_i(\textbf{q}_\ell|\xi) \in [0,1] \; \mathrm{for} \; i \in \llbracket 0,N_\mathrm{T}-1 \rrbracket; \; \sum_{i=0}^{N_\mathrm{T}-1} \mathcal{T}_i(\textbf{q}_\ell|\xi) = 1 .
\end{equation} 
Here, $\mathcal{T}_i(\textbf{q}_\ell|\xi) \equiv \left\langle \mathrm{T}_i(\textbf{q}_\ell|\xi) \right\rangle_{\mathpzc{P}(\mathrm{T}_i(\textbf{q}_\ell)|d,\xi)} = \mathpzc{P}(\mathrm{T}_i(\textbf{q}_\ell)|d,\xi)$ are the posterior probability distribution functions (pdfs) for particle $\textbf{q}_\ell$ to belong to structure $\mathrm{T}_i$, as defined by classifier $\xi$, conditional on the data $d$. These are estimated by counting the relative frequencies at each individual particle position within the set of $\xi$-realizations (see also \citealp{Jasche2010a}; \citetalias{Leclercq2015ST}). Therefore, the cosmic web-type pdf mean on the Lagrangian grid is given by
\begin{equation}
\left\langle \mathpzc{P}(\mathrm{T}_i(\textbf{q}_\ell)|d,\xi) \right\rangle = \frac{1}{N}\sum_{n=1}^{N} \sum_{j=0}^{N_\mathrm{T}-1} \updelta_{\mathrm{K}}^{\mathrm{T}_i(\textbf{q}_\ell),\mathrm{T}^n_j(\textbf{q}_\ell|\xi)} ,
\label{eq:pdf_mean}
\end{equation}
where $n$ labels one of the $N$ samples, $\mathrm{T}^n_j(\textbf{q}_\ell|\xi)$ is the result of classifier $\xi$ on the $n$-th sample at particle position $\textbf{q}_\ell$ (i.e. a unit $N_\mathrm{T}$-vector containing zeros except for one component, which indicates the structure type), and $\updelta_{\mathrm{K}}$ is a Kronecker symbol.

Using the above definitions, we obtain probabilistic maps of structures on the Lagrangian grid describing the early Universe in the volume probed by SDSS main sample galaxies. More precisely, for each classifier $\xi$, we obtain a probability mass function (pmf) at each particle position on the Lagrangian grid, indicating the possibility to encounter a specific structure type at that position. These pmfs consist of $N_\mathrm{T}$ numbers in the range $[0,1]$, $\mathpzc{P}(\mathrm{T}_i(\textbf{q}_\ell)|d,\xi)$, that sum up to one for each particle. They are approximated by the mean of each structure pdf (see equation \eqref{eq:pdf_mean}).

In figure \ref{fig:pdf_initial}, we show slices through these web-type posterior probabilities. The rows represent the structures inferred using {\diva}, {\lich}, and {\origami}, respectively. For comparison, the reader is referred to figure 6 in \citetalias{Leclercq2015ST}, which shows structures inferred in the initial density field using the {\tweb}. Note that the present results are defined on the Lagrangian grid of $512^3$ particles, whereas the {\tweb} result is defined on the $256^3$-voxel grid on which initial density fields are discretized. The plot shows the anticipated behavior, with values close to certainty (i.e. zero or one) in regions covered by data, while the unobserved regions (at high redshift or out of the survey boundary) approach a uniform value corresponding to the prior. It is worth noting that different classifiers can have very different prior probabilities for the same structure types \citep[these are given in table II in][]{Leclercq2016CIT}. The good agreement between {\diva} and {\lich} at the level of individual samples, coming from the absence of a substantial vector part in $\Psi$, is also correctly propagated to the posterior mean.

A voxel-wise quantification of the information gain due to SDSS galaxies is provided by the relative entropy of corresponding posteriors and priors (\citetalias{Leclercq2015ST}; \citealp{Leclercq2016CIT}). We applied this technique and found that the Lagrangian web-type posteriors presented in figure \ref{fig:pdf_initial} yield much higher information gain than the application of the {\tweb} to initial density fields. Averaged over the constrained volume, this number is $\sim 0.404$~Sh for {\diva}, $\sim 0.297$~Sh for {\lich}, and $\sim 0.462$~Sh for {\origami}, as compared to $\sim 0.196$~Sh for the {\tweb}. This consideration substantiates the non-local transport of information along Lagrangian trajectories operated by the structure formation model, and demonstrates that a probabilistic description of the Lagrangian dark matter sheet yields a much higher information content for structure type classification than previous methods.

\section{Consequences in Eulerian coordinates}
\label{sec:Consequences in Eulerian coordinates}

In this section, we describe potentially observable consequences of our inference of the nearby Lagrangian dark matter phase-space sheet. To do so, we analyze the properties of our constrained simulations in Eulerian coordinates in light of previous Lagrangian results. In other words, we demonstrate that our analysis of dark matter phase-space properties has an interesting degree of predictive power.

Formally, the problem can be expressed as follows. Before the data are considered, the distribution of an unknown observable $y$ is its \textit{prior predictive distribution},
\begin{equation}
\mathpzc{P}(y) = \int \mathpzc{P}(y,\delta^\mathrm{i}) \, \mathrm{d}\delta^\mathrm{i} = \int \mathpzc{P}(\delta^\mathrm{i}) \, \mathpzc{P}(y|\delta^\mathrm{i}) \, \mathrm{d}\delta^\mathrm{i}
\end{equation}
where $\delta^\mathrm{i}$ are the three-dimensional initial conditions. After the data $d$ have been observed, the distribution for $y$ is the \textit{posterior predictive distribution},
\begin{eqnarray}
\mathpzc{P}(y|d) & = & \int \mathpzc{P}(y,\delta^\mathrm{i}|d) \, \mathrm{d}\delta^\mathrm{i} \nonumber \\
& = & \int \mathpzc{P}(\delta^\mathrm{i}|d) \, \mathpzc{P}(y|\delta^\mathrm{i},d) \, \mathrm{d}\delta^\mathrm{i} \nonumber \\
& \approx & \frac{1}{N} \sum_{n=1}^N \int \updelta_\mathrm{D}(\delta^\mathrm{i}-\delta^\mathrm{i}_n) \, \updelta_\mathrm{D}\!\left(y-\mathcal{Y}(\delta^\mathrm{i})\right) \mathrm{d} \delta^\mathrm{i} \nonumber\\
& \approx & \frac{1}{N} \sum_{n=1}^N \updelta_\mathrm{D}\!\left(y-\mathcal{Y}(\delta^\mathrm{i}_n)\right).
\end{eqnarray}
In the above equations, we have used the {\borg} inference of initial conditions, i.e. $\mathpzc{P}(\delta^\mathrm{i}|d) \approx \sum_n \updelta_\mathrm{D}(\delta^\mathrm{i}-\delta^\mathrm{i}_n)$ where $\delta^\mathrm{i}_n$ is one of the $N$ samples. We have also assumed that $y$ does not further depend on the data once the initial conditions are known, and that $y$ is a deterministic function $\mathcal{Y}$ of the initial conditions, i.e. $\mathpzc{P}(y|\delta^\mathrm{i},d) = \mathpzc{P}(y|\delta^\mathrm{i}) = \updelta_\mathrm{D}\!\left(y-\mathcal{Y}(\delta^\mathrm{i})\right)$. Therefore, under the assumption that inferred initial conditions contain all the information on $y$, our procedure generates samples $y_n \equiv \mathcal{Y}(\delta^\mathrm{i}_n)$ of $\mathpzc{P}(y|d)$.

In the following, we start by looking at phase-space properties traced at Eulerian coordinates, which yields improved estimators for density fields (section \ref{sec:Phase-space properties traced at Eulerian coordinates}), then we show how to translate Lagrangian classifications to Eulerian coordinates (section \ref{sec:Translation of Lagrangian classifications to Eulerian coordinates}), discuss the properties of cosmic velocity fields (section \ref{sec:Cosmic velocity fields in the SDSS volume}), and finally present some cosmographic results (section \ref{sec:Cosmography}). For all results, we also show that observational uncertainties are propagated in a non-linear and non-Gaussian fashion by the {\cola} forward model, from which we get a feeling for how precise the predictions are.

\subsection{Phase-space properties traced at Eulerian coordinates}
\label{sec:Phase-space properties traced at Eulerian coordinates}

\subsubsection{Eulerian secondary stream density}
\label{sec:Eulerian secondary stream density}

\begin{figure*}
\begin{center}
\includegraphics[width=0.70\textwidth]{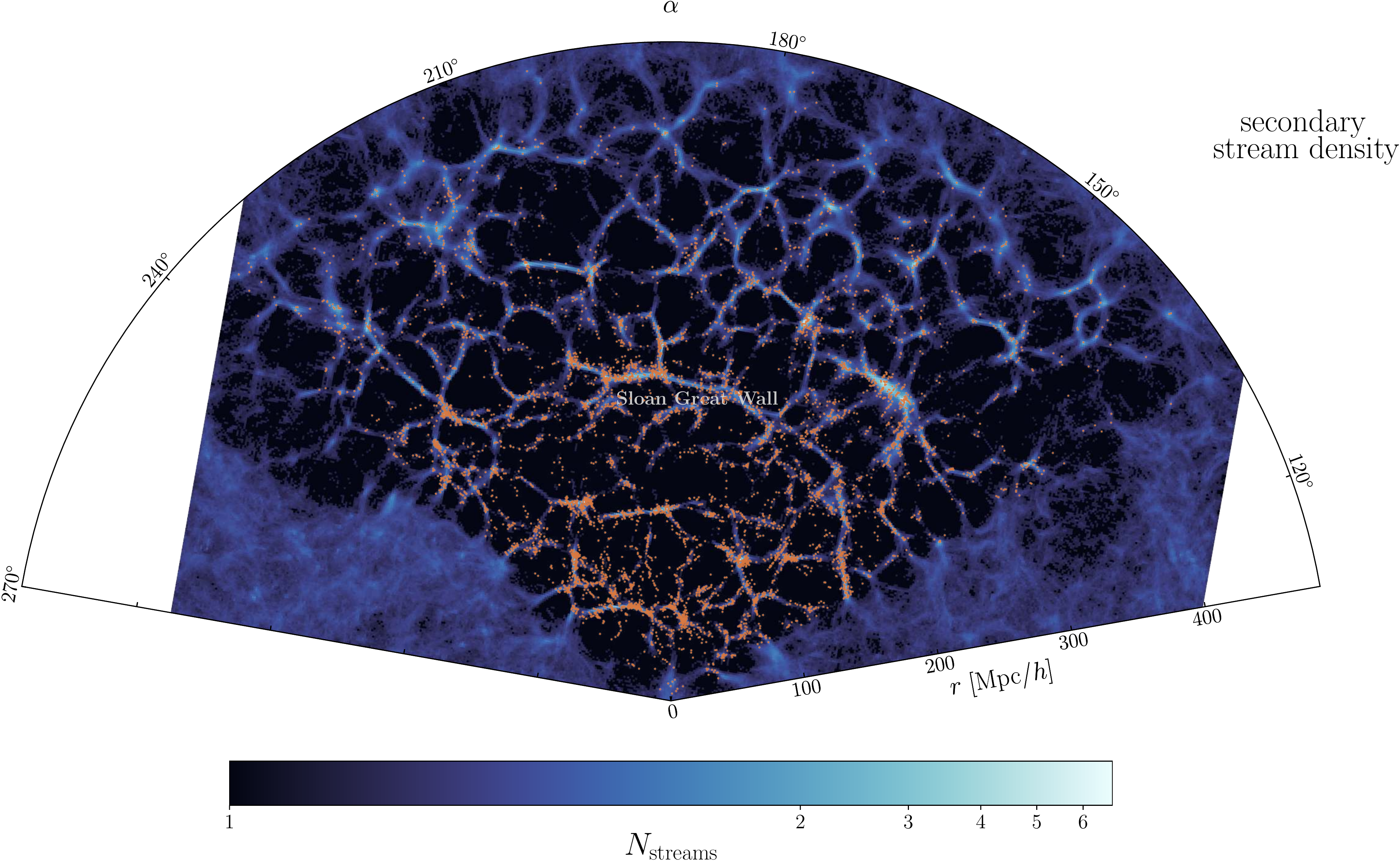}
\caption{Slice through the posterior mean for the Eulerian secondary stream density, defined on a cubic grid of $512^3$ voxels. Galaxies of the SDSS main galaxy sample are overplotted. The slice is $\sim 3$~Mpc/$h$ thick around the celestial equator.\label{fig:secondary_stream_density}}
\end{center}
\end{figure*}

The simplest phase-space property that can be translated into Eulerian coordinates is the number of matter streams. As discussed in section \ref{sec:Tessellation of the Lagrangian grid}, this is simply achieved by counting, in the final configuration of the tessellated Lagrangian dark matter sheet, the number of tetrahedra that contain any given position. The resulting field is called the secondary stream density. Since the dark matter phase-space sheet is discretized using a finite number of particles (and hence of tetrahedra), the secondary stream density is only a lower bound on the actual number of dark matter streams at any single location.

We applied this estimator to our full set of constrained realizations. Figure \ref{fig:secondary_stream_density} shows the resulting posterior mean for the Eulerian secondary stream density. A slice of $\sim$~3~Mpc/$h$ (one voxel) thickness around the celestial equator is shown, and the galaxies of the northern cap of the SDSS main galaxy sample that are contained in the same volume are overplotted.\footnote{The transformation between our set of Cartesian coordinates and equatorial J2000.0 spherical coordinates can be found in appendix B of \citetalias{Jasche2015BORGSDSS}.}

As expected from simulations \citep[e.g.][]{Ramachandra2015}, the number of streams correlates well with the galaxy density field: regions of high galaxy density are the ones where several matter flows contribute. For example, in the inner regions of the Sloan Great Wall, which is clearly visible in the middle of the slice, at least 6 matter streams are converging. At large distances ($r \gtrsim 300$~Mpc/$h$), comparing the posterior mean for the Eulerian secondary stream density and galaxy number counts reveals that the predictions remain accurate even in regions poorly sampled by galaxies. This comes from the capacity of the {\borg} algorithm to accurately recover structures even in regimes of low signal-to-noise \citepalias[see the discussion in][]{Jasche2015BORGSDSS}.

Our result readily provides a probability distribution for the number of large-scale matter streams at any galaxy position. This information, which is not available from raw survey data, is of potential interest for various cosmological and astrophysical analyses. It is expected to be crucially linked to the formation history of galaxies. If dark matter annihilates, then clear signals are predicted to come from nearby rich galaxy clusters \citep{Gao2012}. In fact, if dark matter is absolutely cold, its mass density (and therefore the annihilation luminosity) should be infinite at caustics \citep{White2009}. Galaxies living in stream-crossed regions are also the ones for which the ``reconstruction'' procedure used to correct for large-scale flows in baryon acoustic oscillations analyses \citep{Eisenstein2007,Padmanabhan2012,Burden2015} will fail, as it relies on the single-stream assumption and on the Zel'dovich approximation.

\subsubsection{Density estimators}
\label{sec:Density estimators}

\begin{figure*}
\begin{center}
\includegraphics[width=\textwidth]{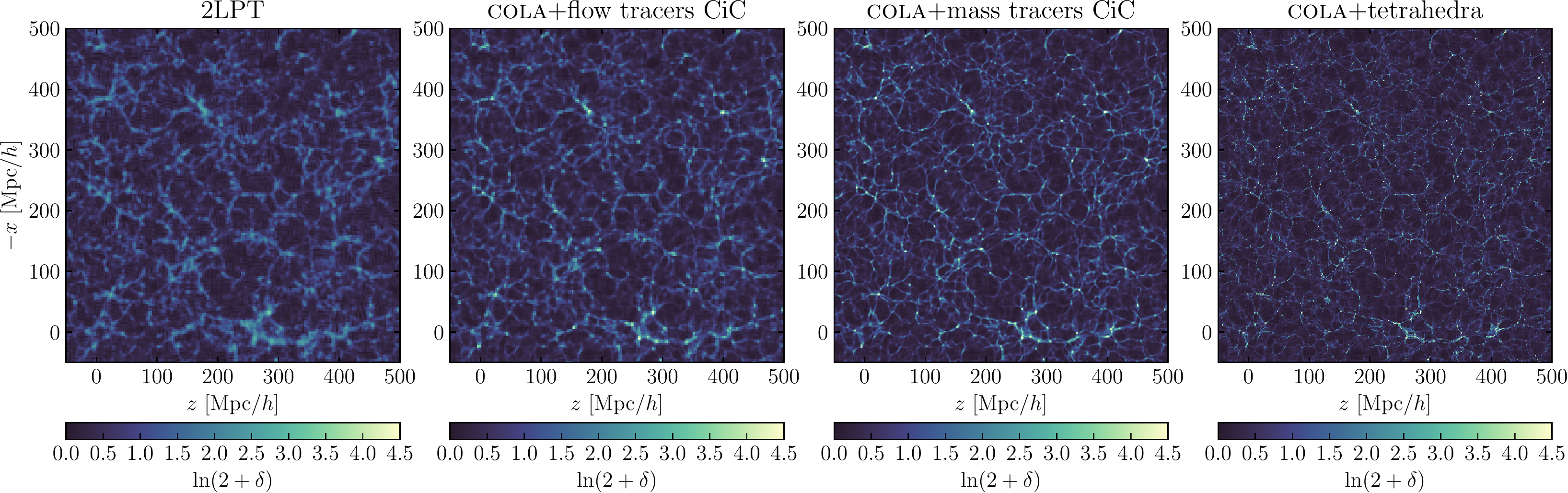}\vspace{4pt}
\includegraphics[width=\textwidth]{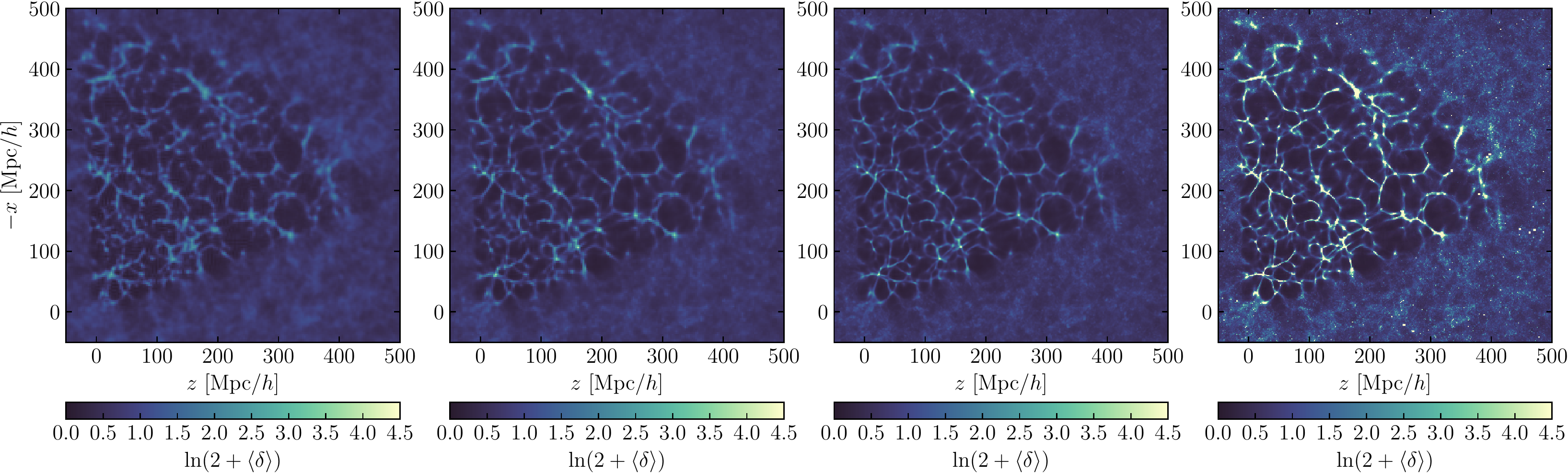}\vspace{4pt}
\includegraphics[width=\textwidth]{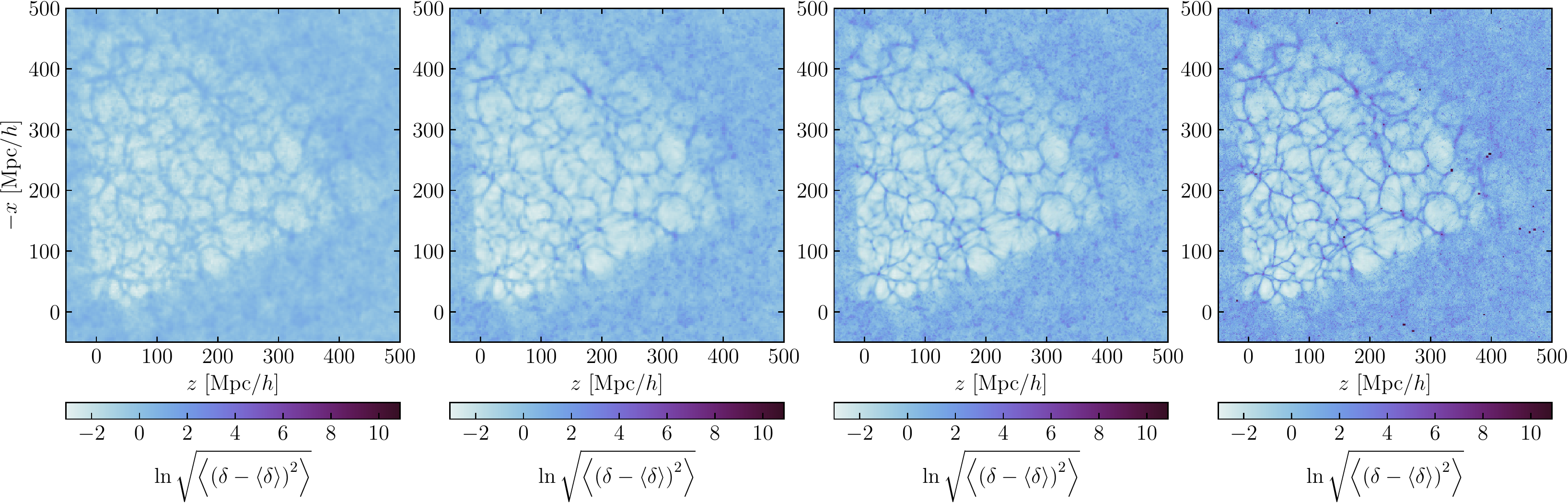}
\caption{The influence of different structure formation models and density estimators on {\borg} samples. Slices through the final density field in one sample (first row), the posterior mean (second row) and the posterior standard deviation (third row) are shown. From left to right, the columns use: 2LPT and CiC of flow tracers (as {\borg}), {\cola} and CiC of flow tracers, {\cola} and CiC of mass tracers, {\cola} and explicit projection of all tetrahedra of the tessellated Lagrangian grid. For the last column only and for visualization purposes, densities are clipped at 800 times the mean density. The improvement of the density estimator (third and fourth columns) constitute an extension of the non-linear filtering procedure (going from the first to the second column).\label{fig:density}}
\end{center}
\end{figure*}

As shown by \citet{Abel2012} and \citet{Hahn2013} in simulations, instead of simply assigning dark matter particles to the grid via CiC deposit, knowledge of their initial positions can be exploited to give better density estimators. Specifically, the tessellated dark matter phase-space sheet yields a density estimator when the four vertices of each tetrahedron are given a weight of $\rho_\mathrm{tet}$ (see section \ref{sec:Tessellation of the Lagrangian grid}). As a simpler alternative, tetrahedral mass elements can be approximated by pseudo-particles (the mass tracers) which are assigned to the grid by standard CiC.

We applied these two new density estimators to our set of constrained simulations. The results are shown in figure \ref{fig:density} in comparison to earlier results. There, the first column represents the raw {\borg} output, for which the structure formation model is 2LPT \citepalias{Jasche2015BORGSDSS}. The second column represents the standard CiC assignment of usual dark matter particles (flow tracers) in our {\borg}-{\cola} realizations, as obtained in \citetalias{Leclercq2015ST}. The third and fourth columns are the new density estimators: the third raw corresponds to the CiC assignment of mass tracers, and the fourth row to the explicit projection of all tetrahedra. For each density estimator, we show one sample (first row), the posterior mean (second row) and the posterior standard deviation (third row). Tetrahedra can be projected to an arbitrarily fine grid. Similarly, there are 24 times more mass tracers than flow tracers, but as argued by \citet{Hahn2013}, the pseudo-particle approximation technique yields more than a factor 24 increase in effective mass resolution, when comparing CiC deposit of flow tracers versus mass tracers. For these reasons, we have been able to double the resolution of the Eulerian grid: the third and fourth columns show a grid of $512^3$ voxels versus $256^3$ voxels for the first and second columns.

In individual samples, the plot shows the anticipated behavior, with visually indistinguishable observed and unobserved regions. This indicates the joint performance of {\borg}, {\cola}, and phase-space estimators to augment unobserved regions with physically and statistically correct information. The ensemble of data-constrained realizations also permits to correctly propagate observational uncertainties. As expected, all mean density fields exhibit highly-detailed structures where data constraints are available, and approach the cosmic mean value in unobserved regions. The standard deviation reveals a high degree of correlation between signal and noise, which is typical of non-linear inference problems. In particular, in {\cola} realizations, the densest clusters which cannot be reproduced by 2LPT correspond to high variance regions. In some small patches away from the well-constrained regions, the density estimated from explicit projection of tetrahedra has high posterior mean and variance. These are numerical artifacts driven by one or a few samples, and due to a bias of the estimator in the densest regions \citep[the linearly-interpolated phase-space sheet no longer tracks the true distribution function there, see][]{Abel2012}.

The two new density estimators extend the non-linear filtering procedure originally introduced by \citet{Leclercq2015DMVOIDS} and reviewed in \citet[][chapter 7]{LeclercqThesis}. Starting from inferred density fields, we go forward in time using an arbitrary structure formation model, $\mathcal{SF}$, combined with a density estimator, $\mathcal{DE}$. The result is a non-linear function, noted $\mathcal{G}_\mathrm{NL} \equiv \mathcal{DE} \circ \mathcal{SF}$, which maps the initial density field $\delta^\mathrm{i}$ into a final density field $\delta^\mathrm{f}_\mathrm{NL}$ at a scale factor $a$:
\begin{equation}
\delta^\mathrm{i} \mapsto \delta^\mathrm{f}_\mathrm{NL} = \mathcal{G}_\mathrm{NL}(\delta^\mathrm{i},a).
\end{equation}
The {\borg} data model uses $\mathcal{SF}=$~2LPT and $\mathcal{DE}=$~\{CiC of flow tracers\} \citep{Jasche2013BORG}. In \citet{Leclercq2015DMVOIDS} and \citetalias{Leclercq2015ST}, we improved the physical model over 2LPT by using respectively $\mathcal{SF}=$~\textsc{Gadget-2} \citep{Springel2005} and $\mathcal{SF}=$~{\cola} \citep{Tassev2013}; but we kept $\mathcal{DE}=$~\{CiC of flow tracers\}. Here, we keep the physical model fixed, $\mathcal{SF}=$~{\cola}, but we improve instead the density estimator, using $\mathcal{DE}=$~\{CiC of mass tracers\} or $\mathcal{DE}=$~\{projection of tetrahedra\}. As shown, this alternative approach also yields improvements of the composite function $\mathcal{G}_\mathrm{NL}$.

\subsection{Translation of Lagrangian classifications to Eulerian coordinates}
\label{sec:Translation of Lagrangian classifications to Eulerian coordinates}

\begin{figure*}
\begin{center}
{\diva}
\includegraphics[width=\textwidth]{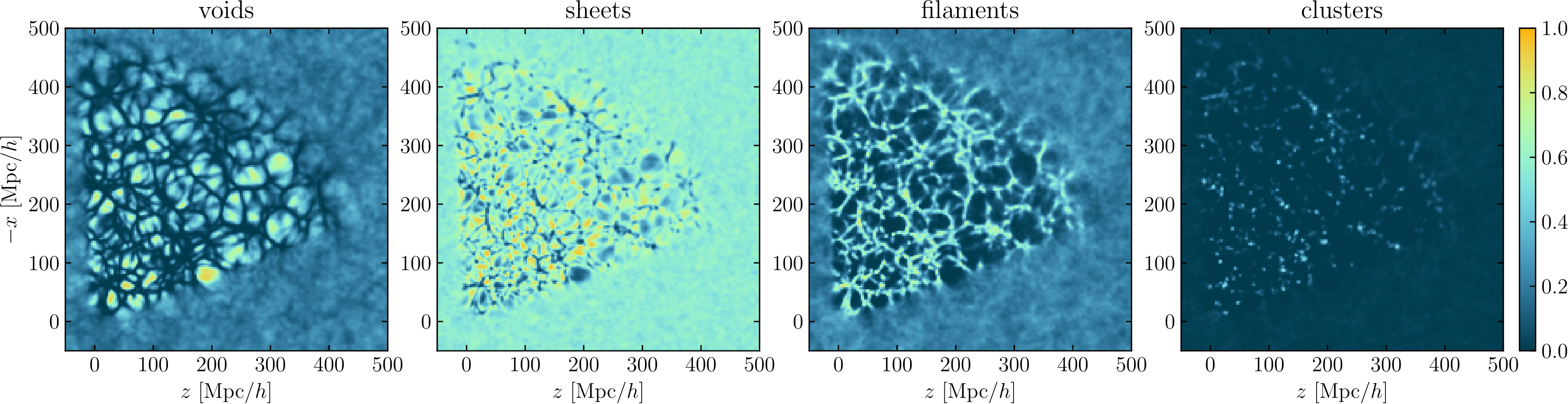}
\rule{4cm}{0.4pt}\\
{\lich}
\includegraphics[width=\textwidth]{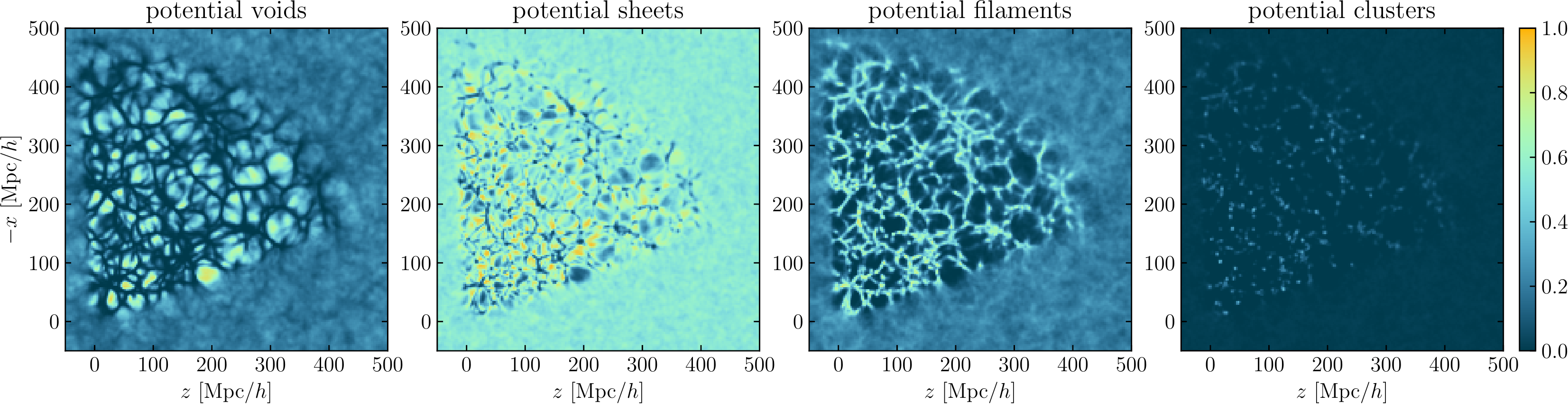}
\includegraphics[width=\textwidth]{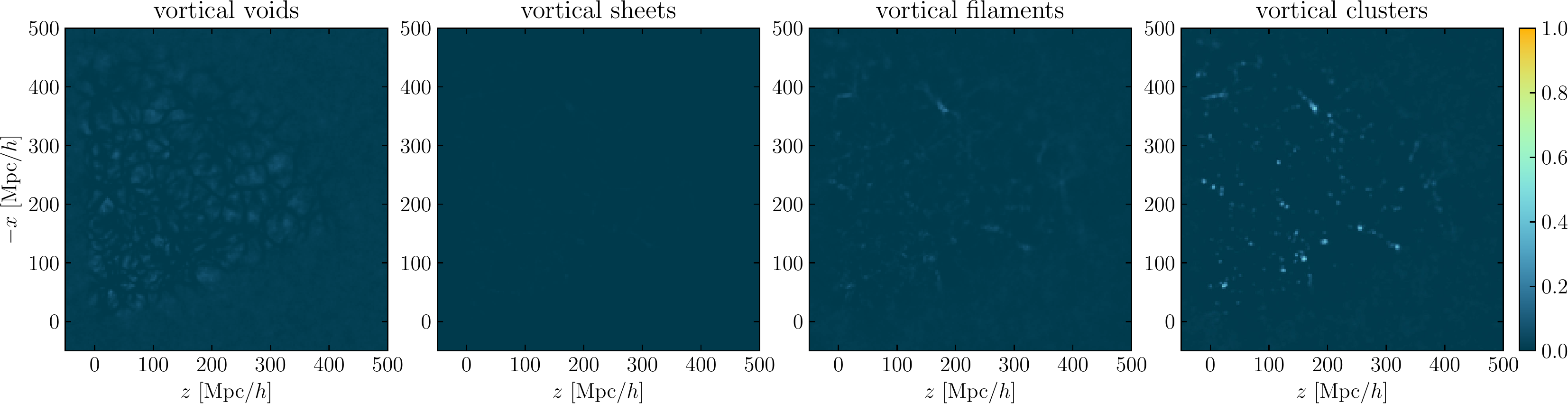}
\rule{4cm}{0.4pt}\\
{\origami}
\includegraphics[width=\textwidth]{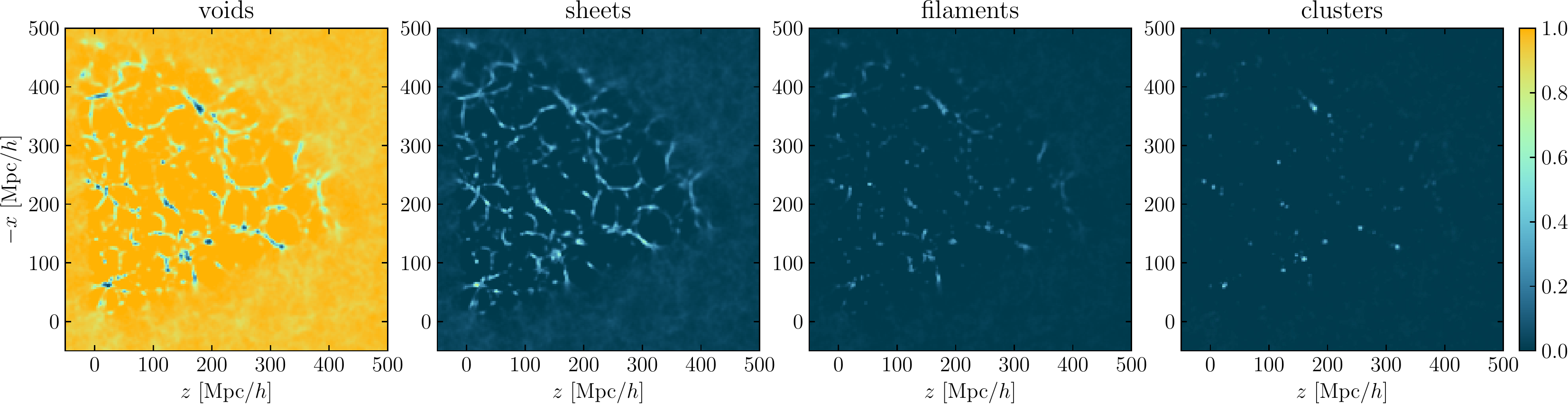}
\caption{Slice through the posterior probabilities for different structure types (from left to right: void, sheet, filament, and cluster), in the late-time large-scale structure in the Sloan volume ($a=1$). Structure types are classified using {\diva} (first row), {\lich} (second and third rows), or {\origami} (fourth row). For {\lich}, potential and vortical structures are distinguished. For each classifier, the four or eight three-dimensional probabilities, defined on a $256^3$-voxel grid, sum up to one on a voxel basis. For comparison, the reader is referred to figure 3 in \citet{Leclercq2015ST}, where structures are classified according to the {\tweb} procedure.\label{fig:pdf_final}}
\end{center}\vspace*{-5pt}
\end{figure*}

In section \ref{sec:Lagrangian classifications}, we have considered classifications of the dark matter sheet, discretized by the simulation particles. In order to classify structures at redshift zero, it is possible to translate the results of Lagrangian classifiers into Eulerian coordinates. As mentioned in section \ref{sec:Tessellation of the Lagrangian grid}, a natural way to do this is to use the tessellation of the Lagrangian lattice into tetrahedra, which allows the translation of any particle-wise quantity to the Eulerian grid. However, since this is rather computationally expensive, we introduce below a simplified procedure to translate Lagrangian classifications into Eulerian coordinates. The latter is less precise at the level of individual samples, but we have checked that it yields negligible differences with the respect to the tessellation approach, as far as the posterior mean (figure \ref{fig:pdf_final}) is concerned.

The simplified approach to translate Lagrangian classifications into Eulerian coordinates works as follows. We consider particles at their final positions and assign them to the Eulerian grid (of $256^3$ voxels in our case) using the CiC scheme. For structure type $\mathrm{T}_i$, each particle gets a weight of $1$ if its Lagrangian classification is $\mathrm{T}_i$, and 0 otherwise. In each voxel, we then count the relative frequencies for each mass-weighted structure type. In this fashion, in each realization $n$, we obtain $N_\mathrm{T}$ probabilities, in the range $[0,1]$, for a final Eulerian voxel to belong to each of the structure types. We denote them by $\mathpzc{P}_n(\mathrm{T}_i(\textbf{x}_k)|d,\xi)$. In general, we have $\mathcal{S}_n(\textbf{x}_k) \equiv \sum_{i=0}^{N_\mathrm{T}-1} \mathpzc{P}_n(\mathrm{T}_i(\textbf{x}_k)|d,\xi) = 1$; this sum is $0$ if and only if, in realization $n$, no particle was assigned to the cell $\textbf{x}_k$ by the CiC scheme (i.e. it is empty as well as its eight neighbors). These voxels are flagged as noisy and discarded.

It is then possible to go from these per-sample pdfs to a global pdf for the entire analysis, in a similar way as before: we count the relative frequencies of the probabilities at each Eulerian spatial position within the set of samples. With this definition, the web-type posterior mean is given by
\begin{equation}
\left\langle \mathpzc{P}(\mathrm{T}_i(\textbf{x}_k)|d,\xi)\right\rangle = \frac{1}{N_{\textbf{x}_k}} \sum_{n=1}^N \mathpzc{P}_n(\mathrm{T}_i(\textbf{x}_k)|d,\xi),
\end{equation}
where $n$ labels one of the $N$ samples, and the normalization $N_{\textbf{x}_k}$ in voxel $\textbf{x}_k$ is
\begin{equation}
N_{\textbf{x}_k} = \sum_{n=1}^N \updelta_{\mathrm{K}}^{\mathcal{S}_n(\textbf{x}_k),1} .
\end{equation}

In figure \ref{fig:pdf_final}, we show slices through the redshift-zero web-type posterior mean for the different structures as defined by {\diva}, {\lich}, and {\origami}. As before, the plot demonstrates that we are able to propagate typical observational uncertainty to structure type classification. Consistently with the results of section \ref{sec:Stream properties}, where we argued for the accuracy of the approximation of $\Psi$ to a potential field, we find only small differences between {\diva} voids, sheets, and filaments, and the corresponding {\lich} potential structures. However, unsurprisingly, {\lich} allows a more physical detection and characterization of clusters, where the potential flow assumption breaks down. At this point, the reader is referred to figure 3 in \citetalias{Leclercq2015ST} for our map of the same volume, inferred using the {\tweb} definition. The {\tweb} and {\diva} maps are visually similar, with an overall smoother structure for the voids defined by {\diva}, which are sharply separated by walls. {\diva} and {\lich} also visually excel in highlighting the filamentary structure of the cosmic web. In contrast, with {\origami}, most of the volume is filled by voids and more complex, shell-crossed structures are rarely identified.

\subsection{Cosmic velocity fields in the SDSS volume}
\label{sec:Cosmic velocity fields in the SDSS volume}

\begin{figure*}
\begin{center}
\includegraphics[width=0.70\textwidth]{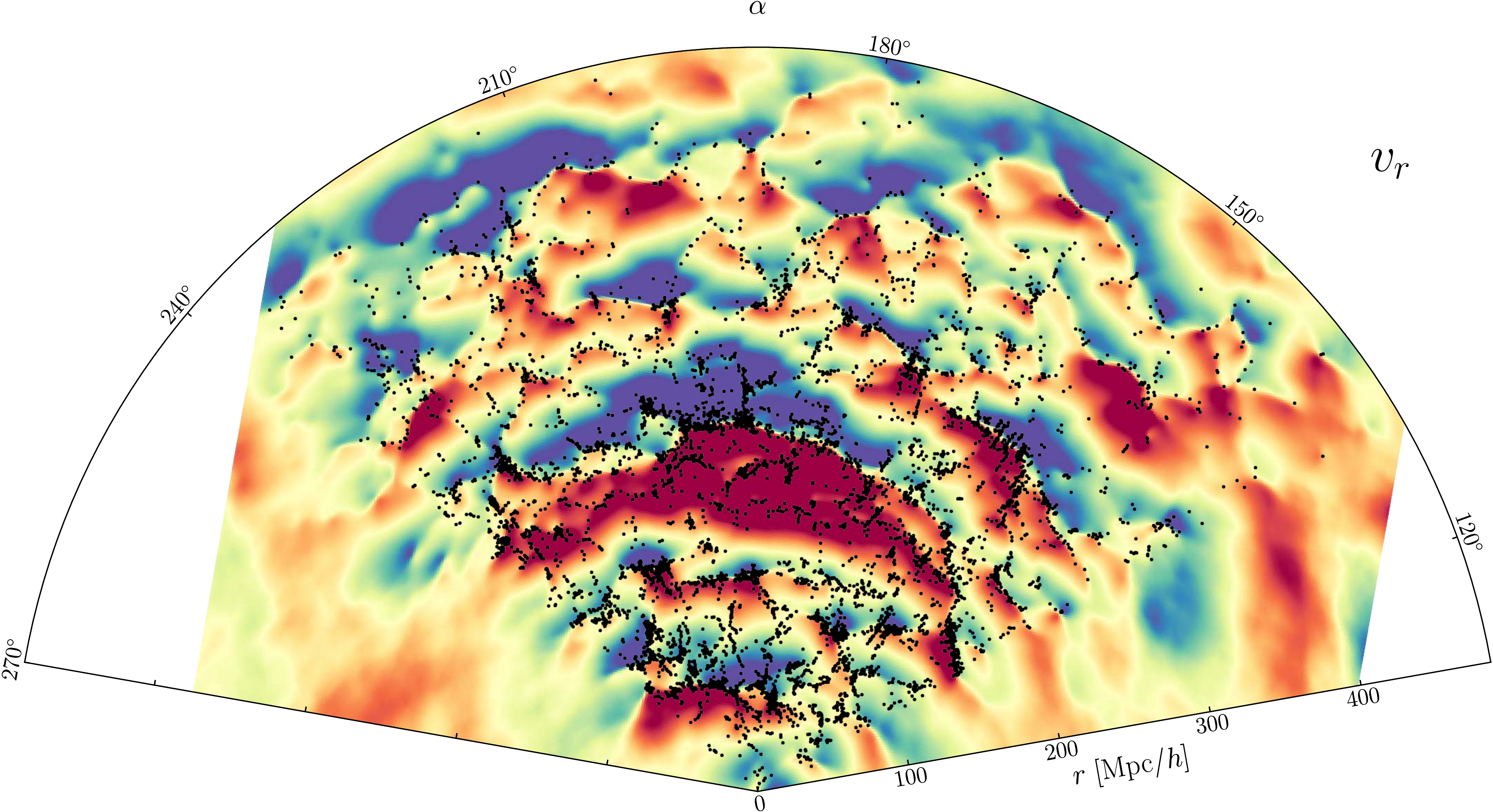}
\includegraphics[width=0.70\textwidth]{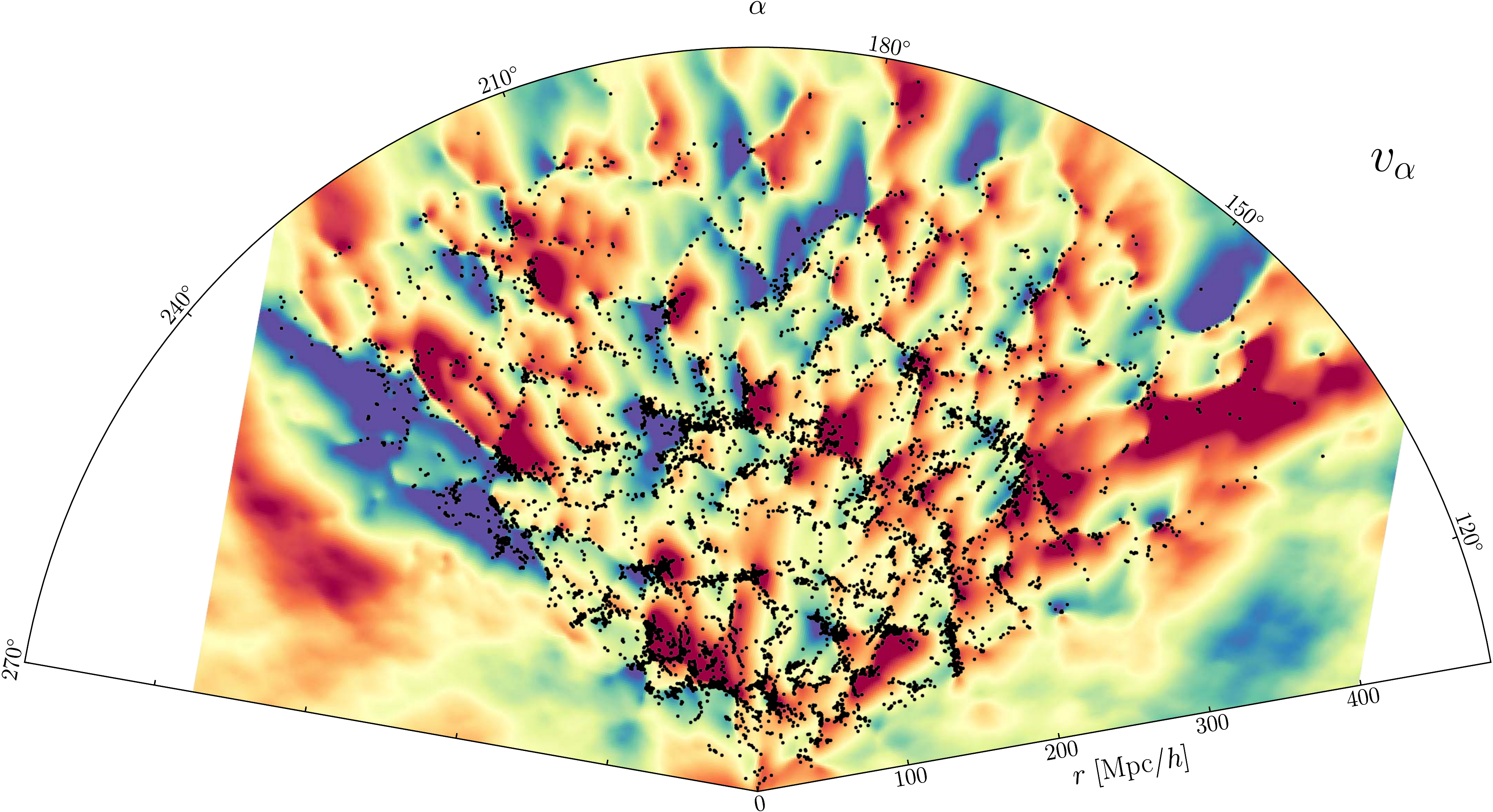}
\includegraphics[width=0.70\textwidth]{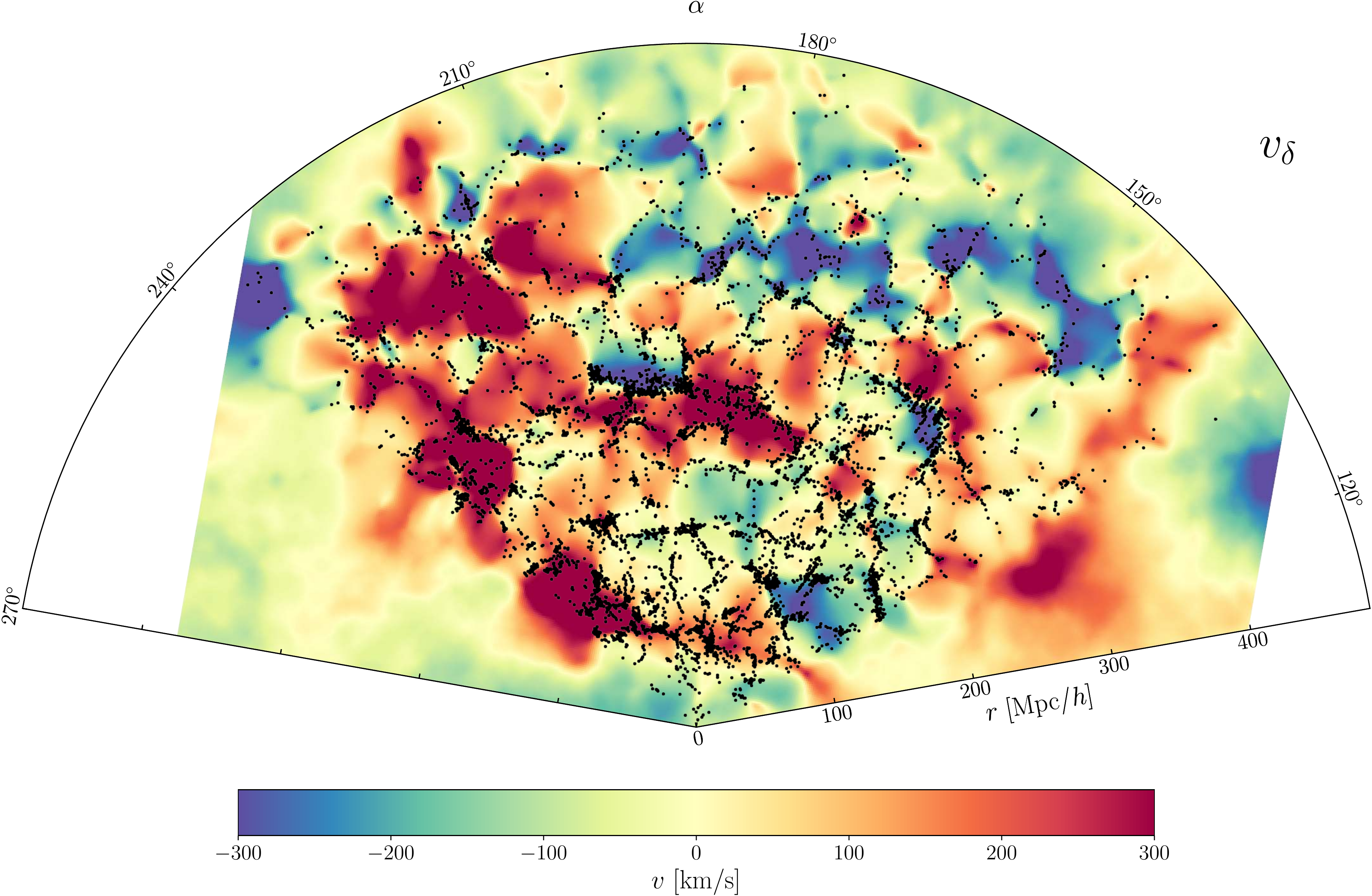}
\caption{Slices through the posterior mean for the radial (upper panel), polar (middle panel) and azimuthal (lower panel) velocity components, defined on a cubic grid of $512^3$ voxels. Velocity fields are obtained from the final configuration of a tetrahedral tessellation of the Lagrangian particle grid. Galaxies of the SDSS main galaxy sample are overplotted. The slice is $\sim 3$~Mpc/$h$ thick around the celestial equator. The reader is invited to compare with figure 7 in \citet{Jasche2015BORGSDSS}.\label{fig:velocity}}
\end{center}
\end{figure*}

\begin{figure}
\begin{center}
\includegraphics[width=\columnwidth]{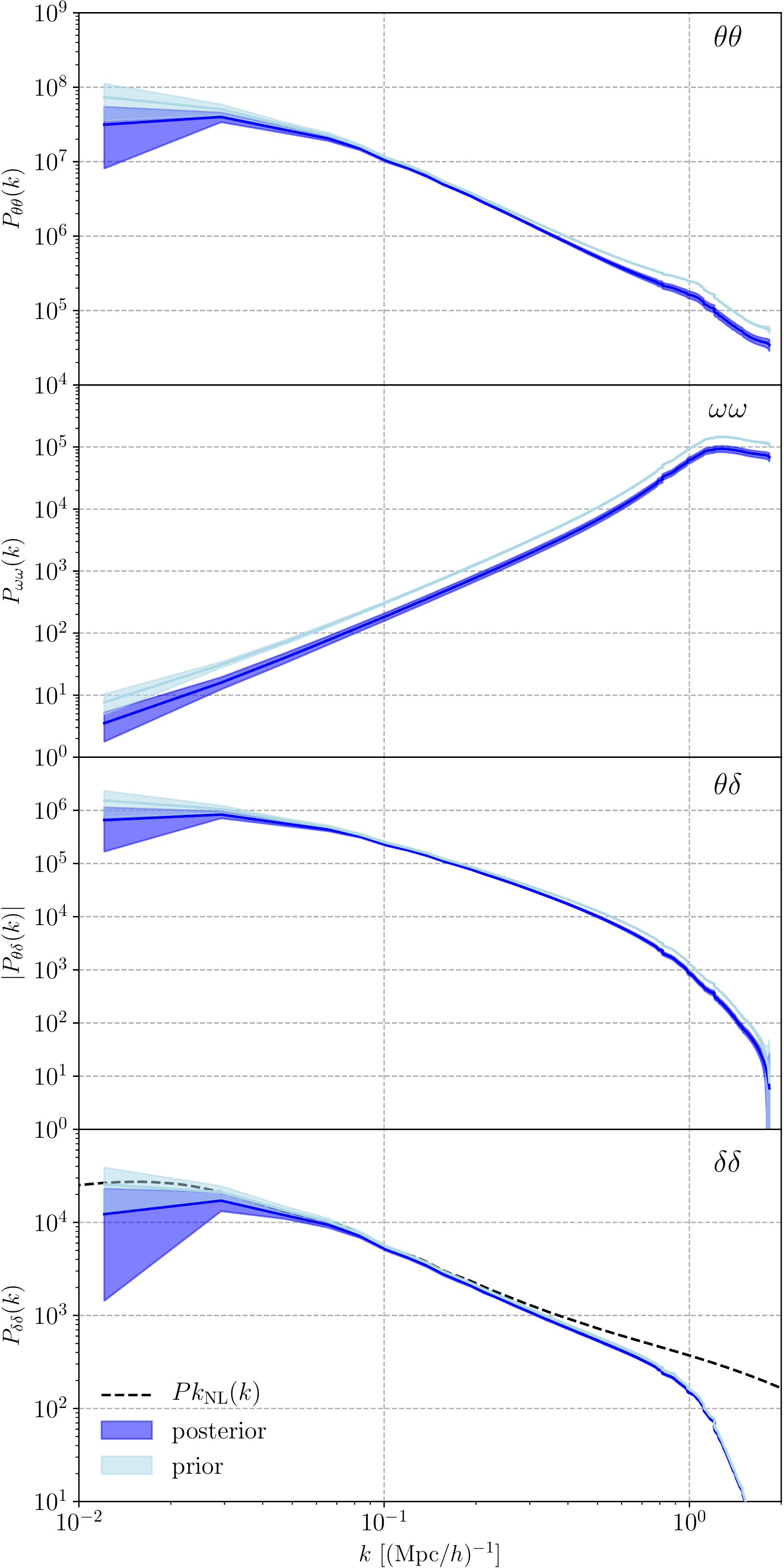}
\caption{Power spectra of the velocity divergence ($\theta\theta$, top panel), the vorticity ($\omega\omega$, second from top), the cross-correlation between velocity divergence and density contrast ($\theta\delta$, third from top) and the density power spectrum ($\delta\delta$, bottom). The particle distributions are determined using constrained (``posterior'') and unconstrained (``prior'') realizations. The solid lines correspond to the mean among all realizations used in this
work, and the shaded regions correspond to the 2-$\sigma$ credible interval estimated from the standard error of the mean. In the bottom panel, the dashed black curve represents $P_\mathrm{NL}(k)$, the theoretical power spectrum expected at $z = 0$ from high-resolution $N$-body simulations. In spite of the approximation made in the inference with 2LPT, {\cola} filtering regenerates most of the final vorticity, so that the power spectrum is only off by 35\% with respect to prior expectations.\label{fig:power_thetaomegadelta}}
\end{center}
\end{figure}

Understanding large-scale velocity fields is of central importance for modern cosmology: it is linked to galaxy and large-scale structure formation \citep{Pichon1999}, is crucial in modeling redshift-space distortions \citep{Percival2009} and in measuring the kinetic Sunyaev-Zel'dovich effect \citep{Shao2011,LavauxAfshordiHudson2013}. Unfortunately, determining the continuous velocity field is an intrinsically complicated problem, even in simulations, because (i) it is only sampled at discrete locations, with poor resolution in low-density regions; and (ii) it is multi-valued in regions where multiple streams contribute. Typical observational uncertainties (mask, selection effects, galaxy biases) add an additional layer of complexity. Recently, \citet{HahnAnguloAbel2015} showed that the first issues can be overcome in simulations by using the Lagrangian phase-space sheet. We apply this technique to our set of constrained simulations and demonstrate that we can also circumvent the observational complications. In doing so, we provide unprecedentedly accurate reconstruction of cosmic velocity fields in the volume of the nearby Universe covered by the northern galactic cap of the SDSS main galaxy sample and its surroundings. 

We applied the estimators described in section \ref{sec:Tessellation of the Lagrangian grid} to our set of constrained simulations, and obtained the components of the velocity field $v_x$, $v_y$, $v_z$, on Eulerian grids of $256^3$ and $512^3$ voxels. As shown in the beginning of section \ref{sec:Consequences in Eulerian coordinates}, under the assumption that velocities are conditionally independent of the data once initial conditions are known, these samples are drawn from $\mathpzc{P}(\textbf{v}|d)$. As before, each of them gives one plausible realization of the velocity field, and the variation between samples quantifies the remaining uncertainty. In figure \ref{fig:velocity}, we show the posterior mean for the three spherical components of the velocity field, in the celestial equator. The interested reader may want to compare with figure 7 in \citetalias{Jasche2015BORGSDSS}, which showed the velocity field as probed by 2LPT particles in one particular sample. The present work constitutes a substantial improvement over these results, since the velocity field is now predicted by a fully non-linear structure formation model instead of 2LPT, well-sampled even in low-density regions, and with more straightforward uncertainty quantification. The recovered velocity field is fully compatible with the known structure of the nearby Universe; showing, for example, the infall of matter onto the Sloan Great Wall. 

As noted in \citetalias{Jasche2015BORGSDSS}, velocity fields are derived \textit{a posteriori} with {\borg}, i.e. they are a prediction of physical models given the inferred initial density field.\footnote{In contrast, algorithms like \textsc{virbius} \citep{Lavaux2016} explicitly exploit velocity information contained in the data to constrain the result, and reconstructions based on surveys of peculiar velocities have produced highly appealing results \citep[e.g.][]{Courtois2013,Tully2014}.} However, since velocity fields are predominantly governed by large-scale modes, which are accurately recovered by {\borg} \citepalias{Jasche2015BORGSDSS}, any improvement in the inference scheme is not expected to crucially change our results. 

To characterize the accuracy at which our velocity fields are predicted, we analyzed their spectral properties against that of reference simulations produced with the same setup \citepalias[presented in][]{Leclercq2015ST}. From the Cartesian velocity components,\footnote{We do not use the divergence and curl directly computed using Lagrangian transport, but rather from the velocity field itself, for the reasons discussed in \citet[][section 5.1]{HahnAnguloAbel2015}.} we computed the divergence and the curl,
\begin{eqnarray}
\theta & \equiv & \nabla_\textbf{x} \cdot \textbf{v},\\
\boldsymbol{\omega} & \equiv & \nabla_\textbf{x} \times \textbf{v},
\end{eqnarray}
using FFTs on the Eulerian grid of side length 750~Mpc/$h$ and $256^3$ voxels. We computed their respective power spectra, given by
\begin{eqnarray}
\left\langle \theta(\textbf{k}) \, \theta(\textbf{k}') \right\rangle & \equiv & P_{\theta\theta}(k) \, \updelta_\mathrm{D}(\textbf{k}-\textbf{k}'),\\
\left\langle \boldsymbol{\omega}(\textbf{k}) \cdot \boldsymbol{\omega}(\textbf{k}') \right\rangle & \equiv & P_{\omega\omega}(k) \, \updelta_\mathrm{D}(\textbf{k}-\textbf{k}') \nonumber\\
& = & \sum_{s\in\{x,y,z\}} P_{\omega_s\omega_s}(k) \, \updelta_\mathrm{D}(\textbf{k}-\textbf{k}').
\end{eqnarray}
To test the proportionality between $\theta$ and the density contrast $\delta$, valid at first order in Lagrangian perturbation theory, we also computed the cross- and auto-power spectra
\begin{eqnarray}
\left\langle \theta(\textbf{k}) \, \delta(\textbf{k}') \right\rangle & \equiv & P_{\theta\delta}(k) \, \updelta_\mathrm{D}(\textbf{k}-\textbf{k}'),\\
\left\langle \delta(\textbf{k}) \, \delta(\textbf{k}') \right\rangle & \equiv & P_{\delta\delta}(k) \, \updelta_\mathrm{D}(\textbf{k}-\textbf{k}').
\end{eqnarray}
In this case, $\delta$ is estimated by standard CiC deposit of dark matter particles to the grid, and we apply a correction factor for aliasing effects and suppression of small-scale power close to the Nyquist wavenumber \citep{Jing2005}.\footnote{We here rely on the CiC scheme and not on the phase-space density fields estimators, following the arguments of \citet[][section 5.2]{HahnAnguloAbel2015}.}

In figure \ref{fig:power_thetaomegadelta}, we show $P_{\theta\theta}$, $P_{\omega\omega}$, $P_{\theta\omega}$ and $P_{\delta\delta}$ as a function of the wavenumber $k$. The measurements from our simulations agree with the results of \citet{HahnAnguloAbel2015}. The plots confirm the agreement between unconstrained and constrained density fields ($\delta\delta$) at all scales (already discussed in \citetalias{Leclercq2015ST}).\footnote{We note that a lack of small scale power with respect to theoretical predictions, for $k \gtrsim 0.5~(\mathrm{Mpc}/h)^{-1}$, is observed in both unconstrained and constrained realizations. This is a gridding artifact due to the finite mesh size used to compute power spectra. This scale corresponds to around one quarter of the Nyquist wavenumber.} Remarkable agreement with simulations can also be observed for $\theta\delta$ and $\theta\theta$ correlations. Deviations can only be observed for $k \gtrsim  0.4~(\mathrm{Mpc}/h)^{-1}$. Below this scale, structure formation becomes notably non-linear, meaning that the 2LPT model used in the {\borg} inference process breaks down, which explains the lack of small-scale power when the constrained initial conditions are run forward with a fully non-linear code. The lack of $\theta\delta$ and $\theta\theta$ correlations does not exceed $\sim$~30\% for $k = 1~(\mathrm{Mpc}/h)^{-1}$. Since the velocity field is curl-free in 2LPT, the vorticity $\boldsymbol{\omega}$ is a unique prediction of the physical model accounting for full gravity. Strikingly, the prediction for $P_{\omega\omega}$ still agrees well with the result of unconstrained simulations, meaning that {\cola} filtering reinstates most of the final vorticity, and does so consistently with the inferred phases (see the $\Upsilon$ panel in figure \ref{fig:dmsheet_one_sample}). Discrepancies are limited to an almost scale-independent negative offset of $\sim$~35\%, which can be explained by known limitations of the {\borg} data model: use of 2LPT instead of fully non-linear gravity, only partial fitting of galaxy bias, absence of redshift-space distortions and lightcone effects.

Summing up this section, at the demonstrated degree of accuracy, our results go beyond the state of the art for the characterization of cosmic velocity fields in the SDSS volume, including its low galaxy-density regions. They include the first quantitative prediction of the vorticity field, and come with the possibility of uncertainty quantification at the level of fields and summary statistics.

\subsection{Cosmography}
\label{sec:Cosmography}

\begin{figure*}
\begin{center}
\includegraphics[width=0.56\textwidth]{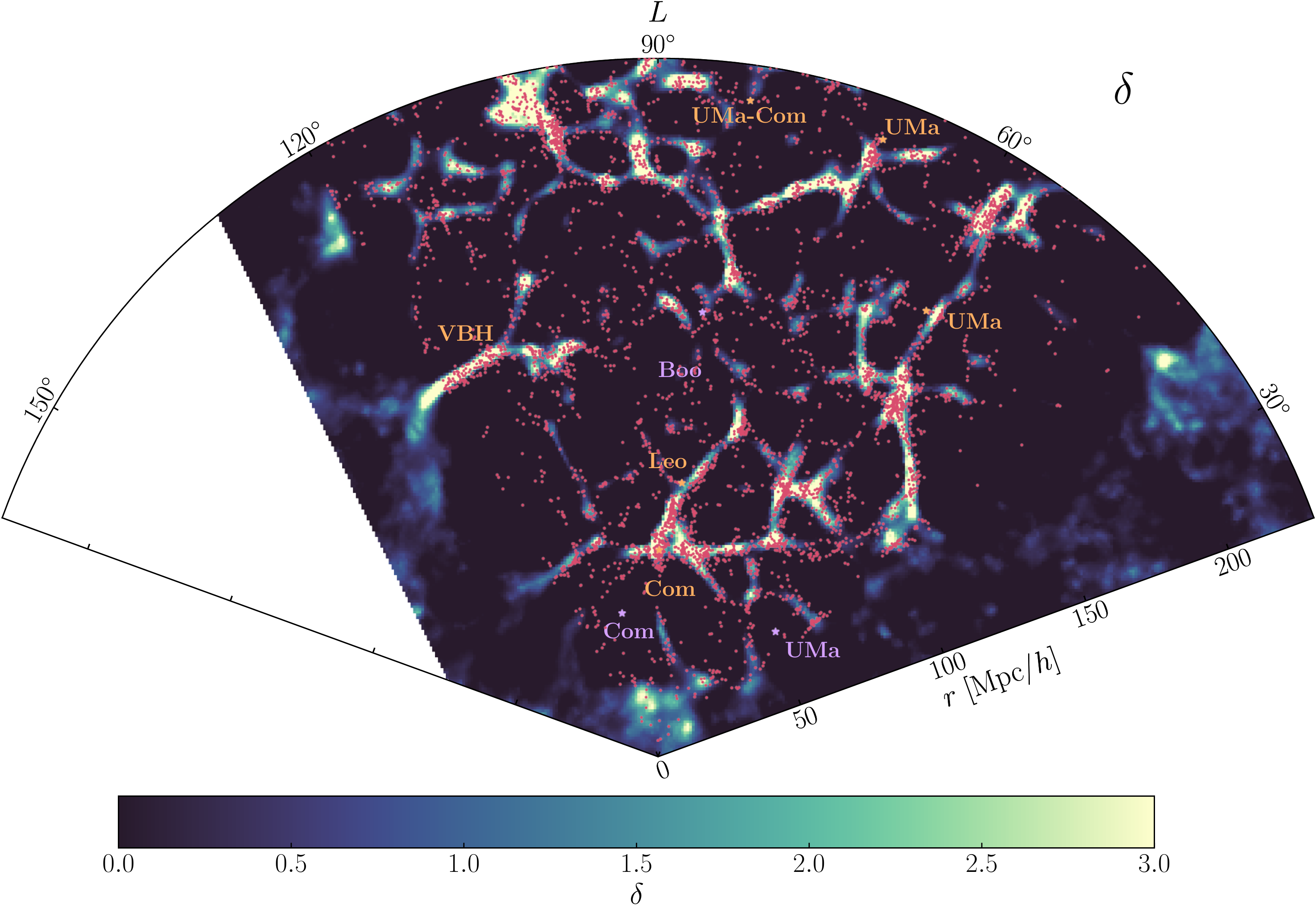}\vspace*{4pt}
\includegraphics[width=0.56\textwidth]{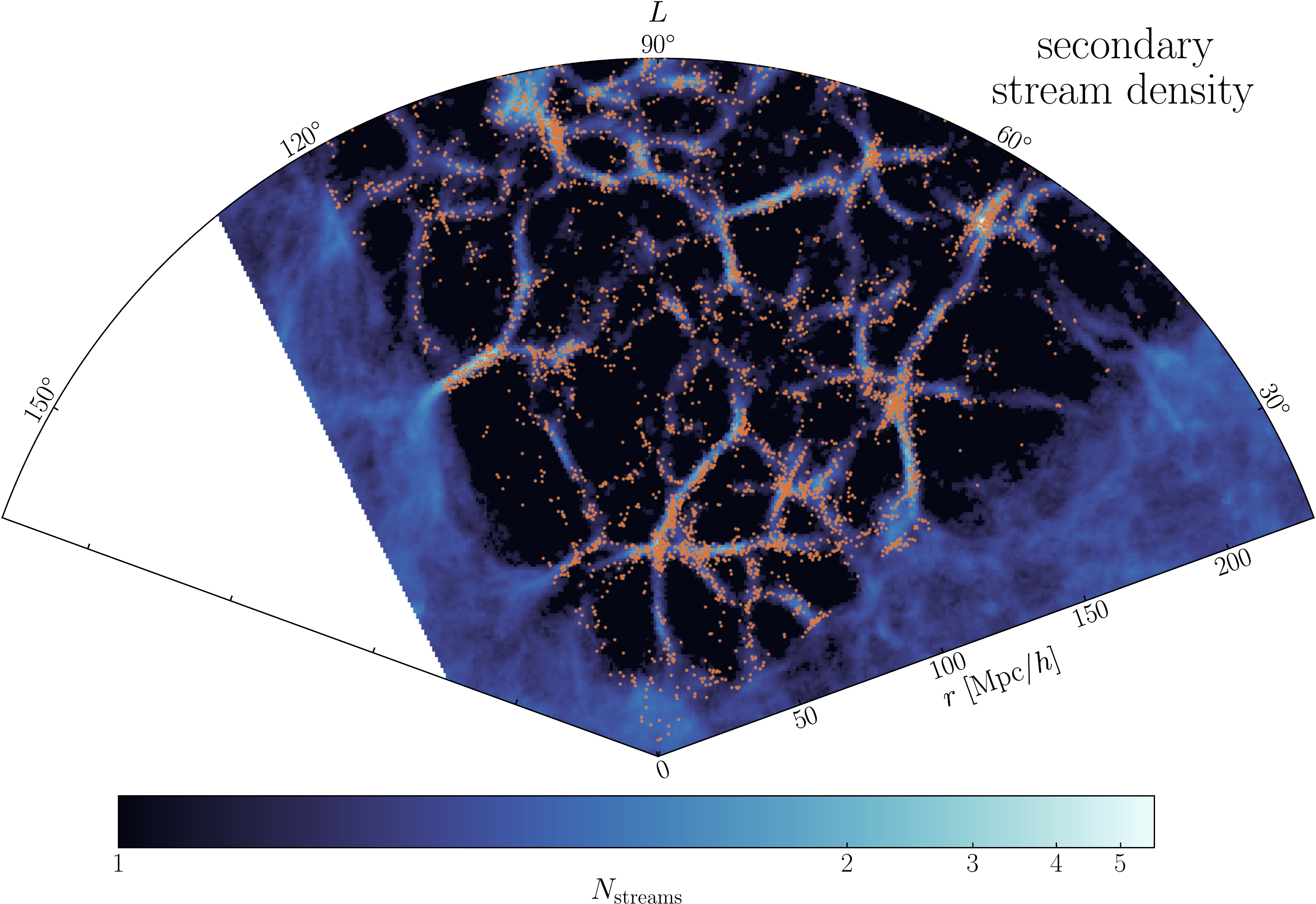}\vspace*{4pt}
\includegraphics[width=0.56\textwidth]{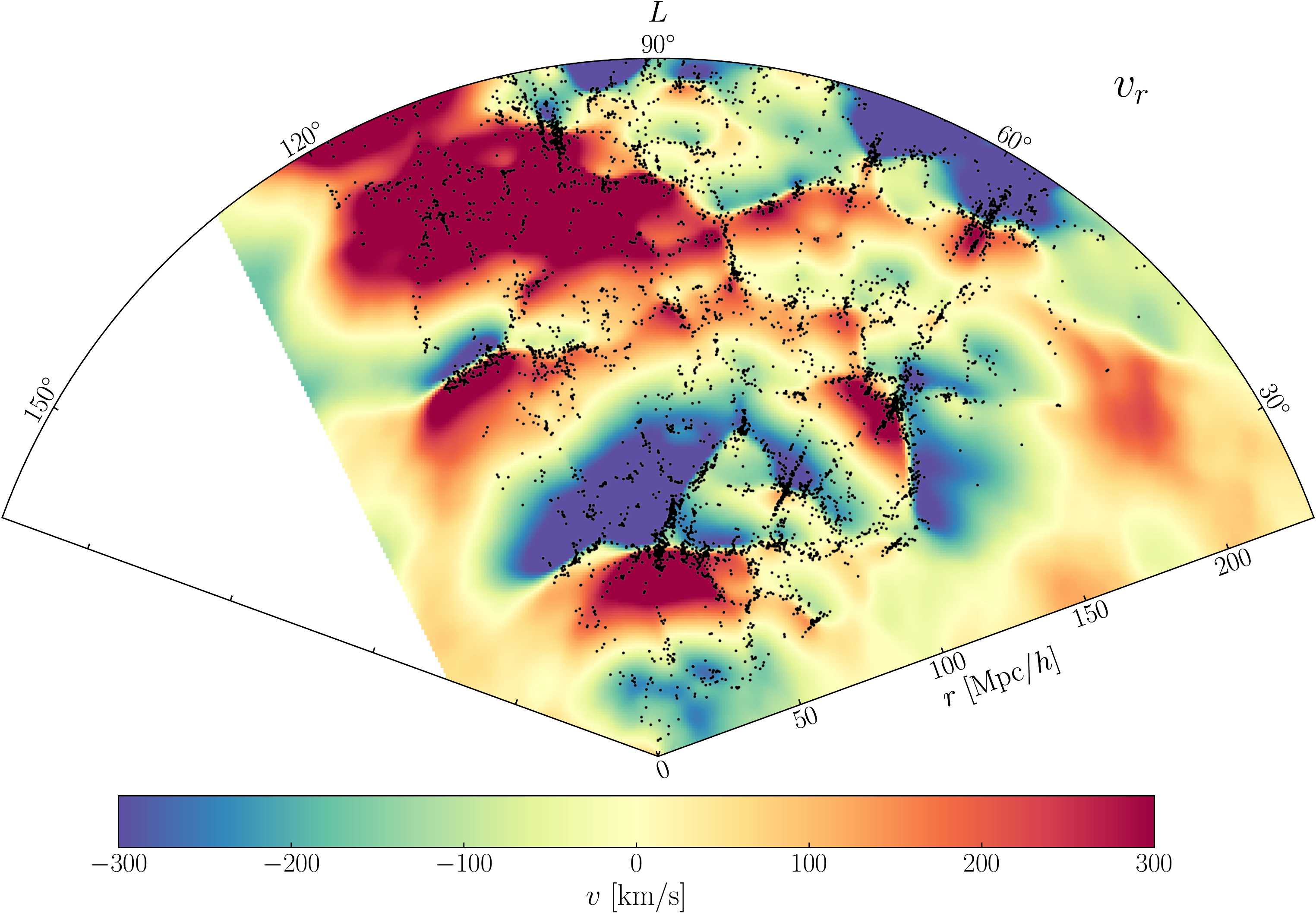}
\caption{Slices through the supergalactic plane, with a thickness of $10$~Mpc/$h$ for galaxies. The fields shown are the density (upper panel), the secondary stream density (middle panel) and the radial velocity (lower panel), as a function of distance and supergalactic longitude $L$. In the upper panel, some structures of the Local Universe are identified in orange: the Coma cluster, the Ursa-Majoris supercluster, and two farther clusters in the direction of Ursa-Major and Ursa-Major/Coma-Berenices. The Virgo-Bo\"{o}tes-Hercules (VBH) filament is also labeled. In purple, two nearby voids in the directions of Ursa-Major and Coma-Berenices, as well as the Bo\"{o}tes void. The centre of superclusters and voids is projected onto the supergalactic plane and denoted as stars.\label{fig:supergalactic}}
\end{center}
\end{figure*}

For illustration purposes, as well as to demonstrate the power of inferring the phase-space structure of dark matter, we discuss in this section some cosmographic results concerning well-known objects of the Local Universe. We have already discussed the Sloan Great Wall \citep{Gott2005,Einasto2011}, at celestial equatorial coordinates $220\degree \lesssim \alpha \lesssim 150\degree$, $-4\degree \lesssim \delta \lesssim 8\degree$, $150 \lesssim r \lesssim 300$~Mpc/$h$, visible in figures \ref{fig:dmsheet_one_sample}, \ref{fig:secondary_stream_density} and \ref{fig:velocity}.

In figure \ref{fig:supergalactic}, we show slices through the supergalactic plane (SGP). The supergalactic coordinate system is defined by the position of its North Pole ($B=90\degree$) in the direction of Galactic coordinates ($l=47.37\degree$; $b=+6.32\degree$, \citealp{Lahav2000}). The portion of the SGP ($B=0\degree$) which is constrained in our analysis corresponds to $30 \lesssim L \lesssim 120\degree$, the rest is not covered by the SDSS main galaxy sample. The different panels show the posterior mean for: the density estimated from mass tracers (see section \ref{sec:Density estimators}), the secondary stream density (see section \ref{sec:Eulerian secondary stream density}), and the radial velocity field (see section \ref{sec:Cosmic velocity fields in the SDSS volume}). Galaxies in a $10$~Mpc/$h$-thick slice are overplotted. Some major structures of the Local Universe are clearly visible: the Coma cluster ($L \sim 90\degree$, $r \sim 70$~Mpc/$h$), the Ursa-Majoris supercluster ($L \sim 60\degree$, $r \sim 170$~Mpc/$h$), and two farther clusters in the direction of Ursa-Major ($L\sim 70\degree$, $r \sim 220$~Mpc/$h$) and Ursa-Major/Coma-Berenices ($L\sim 82\degree$, $r \sim 220$~Mpc/$h$) \citep[e.g.][]{Abell1989}. One can also notice part of Leo supercluster ($L \sim 85\degree$, $r \sim 95$~Mpc/$h$), which is mostly below the SGP (at $B \sim -18\degree$). Three voids intersect the SGP: the well-known Bo\"{o}tes void \citep{Kirshner1981}, whose center is above the SGP ($L \sim 84\degree$, $B \sim 18\degree$, $r \sim 150$~Mpc/$h$) and which is around $200$~Mpc/$h$ in radius \citep{Batuski1985}, a void in the direction of Ursa-Major ($L \sim 47\degree$, $r \sim 60$~Mpc/$h$), and one in the direction of Coma-Berenices ($L \sim 104\degree$, $r \sim 50$~Mpc/$h$) \citep[e.g.][]{Fairall1998}. Part of the Virgo-Bo\"{o}tes-Hercules (VBH) filament, which extends from $L \sim 100\degree$ to $L \sim 150\degree$ at $r \sim 140$~Mpc/$h$ \citep{Lavaux2016BORG2MPP} is also visible. For comparison, see also figure 4 in \citet{Carrick2015} and figure 7 in \citet{Lavaux2016BORG2MPP}.

As expected, structures observed in galaxies correlate well with the reconstructed dark matter density. At this mass resolution, up to 5 matter streams converge in the inner regions of Coma, the Ursa-Majoris supercluster, and the VBH filament. The voids in Coma-Berenices and in Ursa-Major are single-stream regions, as well as most voids of the Local Universe. However, our analysis reveals some substructure of the Bo\"{o}tes void. Reconstructed velocity fields gives further dynamical insight into these structures. Superclusters are still accreting matter, as seen in radial velocity ``dipoles''. The ongoing emptying of voids is also clearly observable; for example, dark matter inside the Bo\"{o}tes void is streaming out towards its surrounding overdensities, the Coma cluster and the VBH filament.

\section{Summary and conclusions}
\label{sec:Summary and conclusions}

This work discusses a fully probabilistic, chrono-cosmographic analysis of the Lagrangian dark matter sheet in a volume covered by the northern cap of the SDSS main galaxy sample. It is a data application of advanced tools, the application of which was so far limited to post-processing simulations and deemed unfeasible in observations. The result is a wealth of new information about the dark matter underlying the observed galaxy distribution. Importantly, all these results are probabilistic, i.e. come with a quantification of uncertainty at the level of maps.

Specifically, we characterized the number of streams and the flow properties in Lagrangian and Eulerian space; we explained how to detect caustics; we improved estimates of the density field and showed that this can be thought of as an extended non-linear filtering; and we obtained extremely accurate reconstructions of the velocity field and its derivatives (velocity divergence and vorticity). We introduced the Lagrangian classifier {\lich}, a generalization of {\diva} to the case of heterogeneous flows. {\lich} is a more precise assignment of web-types --- distinguishing potential and vortical structures --- that we have used to check the accuracy of the potential flow approximation. Using {\diva}, {\lich}, and {\origami}, we presented the first maps of nearby Lagrangian cosmic web elements, and showed how to translate the result into redshift-zero Eulerian maps. With these definitions, identified structures have a direct physical interpretation in terms of the entire history of matter flows toward and inside them.

All the results shown in this paper are predictions based on fusing data constraints imprinted in the inference products (the initial conditions) and physical information expressed in the numerical structure formation model. At various places, we have checked the accuracy of these predictions. All posterior results are consistent with the respective priors, and deviate only in expected ways where the data model used for inference of the initial conditions has known shortcomings. This analysis therefore supports the standard paradigm of large-scale structure formation. If our tests suggest that data models are not yet accurate enough to allow inference of cosmological model parameters, they reinforce the concept underlying {\borg}, i.e. the general picture of including physical structure formation within survey data analysis. In particular, our first prediction of vorticity matches expectations at the $\sim 30$\% level, which is remarkable given that the inference model only relies on the assumption of a potential flow. A measure of the information content of Lagrangian cosmic web maps shows that it is vastly enhanced with respect to Eulerian ones, due to information transport along Lagrangian trajectories. The natural subsequent question of choosing the optimal cosmic web classifier for certain analyses is not addressed in this work; the required information-theoretic framework is introduced in \citet{Leclercq2016CIT}.

This work constitutes the first three-dimensional mapping of the nearby Lagrangian dark matter sheet, as constrained by the SDSS main galaxy sample. The nature of these maps is unprecedented, as they are data-constrained with a high degree of control on uncertainties. The understanding of the phase-space properties of dark matter is a pathway to treating it in a more physical fashion at cosmological scales, beyond the phenomenological description of the $\Lambda$CDM model. In particular, density fields and locations of caustics can be used for dark matter indirect searches in $\gamma$-ray data from the \textit{Fermi} satellite \citep[see][]{Gao2012}; velocity fields can be used to account for peculiar velocity effects in supernova cosmology \citep[see][]{Davis2011} and to characterize the kinetic Sunyaev-Zel'dovich effect \citep[see][]{Shao2011}; tidal and vorticity fields can be used to gain better control on intrinsic alignments in the context of weak lensing \citep{Codis2015} and CMB lensing \citep{LarsenChallinor2016}. In addition to these specific examples, we believe that vast scientific opportunities to use our maps may emerge from the community. These data products, complementing to our earlier release of Eulerian density fields \citepalias{Jasche2015BORGSDSS} and {\tweb} maps \citepalias{Leclercq2015ST}, are provided as on-line material along with this paper. They are available from the first author's website, currently hosted at \href{http://icg.port.ac.uk/~leclercq/data.php}{http://icg.port.ac.uk/$\sim$leclercq/data.php}.

Beyond the specific region probed in this work, our study demonstrates the possibility of mapping the dark matter sheet in the real Universe. This possibility, so far limited to simulations, is uniquely offered by Bayesian physical inference of the initial conditions from large-scale structure surveys, which allows a self-consistent treatment of the necessary aspects: handling of observational effects and description of structure formation. Thus, our result may have important implications for large-scale structure data analyses, as they allow new confrontations of observational data and theoretical models, and therefore novel tests of the standard paradigm of cosmic web formation and evolution.

\textit{Note added:} After submission of this paper, several related pieces of work appeared in the literature. These include a new cosmic web classifier using the eigenvalues of the Hessian of the multistream field \citep{Ramachandra2017}, a new derivation of the theory of caustics \citep{Feldbrugge2017} and a comparison project of different cosmic web algorithms \citep{Libeskind2017}.

\section*{Statement of contribution}

\small{FL implemented {\cola} and the required cosmic web analysis tools, performed the study, produced the maps and wrote the paper. JJ developed {\borg} and lead the SDSS analysis; GL developed {\diva}; JJ and GL contributed to the development of computational tools. BW was involved in the conception and design of Bayesian large-scale structure inference. BW and WP contributed to the interpretation of results. All the authors read and approved the final manuscript.}

\acknowledgments

\small{FL would like to thank Oliver Hahn, Mark Neyrinck and Rien van de Weygaert for discussions that stimulated this analysis. The manuscript benefited from constructive comments and useful suggestions from the reviewer. We are grateful to Bridget Falck and Mark Neyrinck for making publicly available their {\origami} code, which was used in this work. Numerical computations were done on the Sciama High Performance Compute (HPC) cluster which is supported by the ICG, SEPNet and the University of Portsmouth. This work has also made use of the Horizon Cluster hosted by the Institut d'Astrophysique de Paris; we thank St\'ephane Rouberol for running smoothly this cluster for us.}

\small{FL and WP acknowledge funding from the European Research Council through grant 614030, Darksurvey. JJ is partially supported by a Feodor Lynen Fellowship by the Alexander von Humboldt foundation. BW acknowledges funding from an ANR Chaire d'Excellence (ANR-10-CEXC-004-01) and the UPMC Chaire Internationale in Theoretical Cosmology. This work has been done within the Labex \href{http://ilp.upmc.fr/}{Institut Lagrange de Paris} (reference ANR-10-LABX-63) part of the Idex SUPER, and received financial state aid managed by the Agence Nationale de la Recherche, as part of the programme Investissements d'avenir under the reference ANR-11-IDEX-0004-02. This research was supported by the DFG cluster of excellence ``\href{www.universe-cluster.de}{Origin and Structure of the Universe}''.}

\section*{References}
\bibliography{/home/leclercq/workspace/biblio}

\end{document}